\documentclass[final,5p,times,usenatbib,twocolumn]{elsarticle}
\usepackage{amsmath}
\usepackage[normalem]{ulem}
\usepackage{color}
\usepackage{colortbl}
\usepackage[pdftex,             % Create hyperlinks
            breaklinks=true,%
            colorlinks=true,%
            citecolor=black,
            pdfauthor={Bonnivard et al.},%
            pdftitle={CLUMPY: second release}%
           ]{hyperref}

\newcommand{\gr}{$\gamma{\rm -ray}$}
\newcommand{\grs}{$\gamma$-rays}

\newcommand{\beq}{\begin{equation}}
\newcommand{\eeq}{\end{equation}}
\newcommand{\ud}{\mathrm{d}}
\newcommand{\lcdm}{{\ifmmode \Lambda{\rm CDM} \else $\Lambda{\rm CDM}$\fi}}
\newcommand{\Msol}{\rm{M}_\odot}
\newcommand{\Rsol}{\rm{R}_\odot}
\newcommand{\rhosol}{\rho_\odot}

\newcommand{\Ntot}{N_{\rm tot}}
\newcommand{\alpham}{\alpha_M}
\newcommand{\Rvir}{\rm{R}_{\rm vir}}

\newcommand{\clumpy}{{\tt CLUMPY}}
\newcommand{\cfitsio}{{\tt cfitsio}}
\newcommand{\healpix}{{\tt HEALPix}}
\newcommand{\great}{{\tt GreAT}}
\newcommand{\rootcern}{{\tt ROOT}}
\newcommand{\doxygen}{{\tt Doxygen}}
\newcommand{\fits}{{\tt fits}}
\newcommand{\gsl}{{\tt GSL}}

\definecolor{lightgray}{gray}{0.92}

\journal{Computer Physics Communications}

\begin{document}

\begin{frontmatter}


%%\title{Title\tnoteref{label1}}

\title{\clumpy{}: Jeans analysis, $\gamma$-ray and $\nu$ fluxes from dark matter (sub-)structures}

\author[label1]{Vincent Bonnivard}
\ead{bonnivard@lpsc.in2p3.fr}
\author[label2]{Moritz H\"utten}
\ead{moritz.huetten@desy.de}
\author[label3]{Emmanuel Nezri}
\ead{Emmanuel.Nezri@lam.fr}
\author[label4]{\\Ald\'ee Charbonnier}
%\ead{charbonnier@astro.ufrj.br}
\author[label1]{C\'eline Combet}
\ead{celine.combet@lpsc.in2p3.fr}
\author[label1]{David Maurin}
\ead{dmaurin@lpsc.in2p3.fr}

\address[label1]{Laboratoire de Physique Subatomique et de Cosmologie, Universit\'e Grenoble-Alpes,
CNRS/IN2P3,
53 avenue des Martyrs, 38026 Grenoble, France}
\address[label2]{Humboldt-Universit\"{a}t zu Berlin \&  DESY, Platanenallee 6, D-15738 Zeuthen, Germany}
\address[label3]{Laboratoire d'Astrophysique de Marseille, Universit\'e d'Aix-Marseille \& CNRS, UMR7326,
38 rue F. Joliot-Curie, 13388 Marseille Cedex 13, France}
\address[label4]{Observat\'{o}rio do Valongo, Universidade Federal do Rio de Janeiro, Leadeira Pedro Ant\'{o}nio, 43, CEP 20080-090, Brasil}

\begin{abstract}
We present an update of the \clumpy{} code for the calculation of the astrophysical $J$-factors (from
dark matter annihilation/decay) for any Galactic or extragalactic dark matter halo including
substructures: halo-to-halo concentration scatter may now be enabled, boost factors can include several levels of
substructures, and triaxiality is a new option for dark matter haloes. This new
version takes advantage of the \cfitsio{} and \healpix{} libraries to propose \fits{} output maps using
the \healpix{} pixelisation scheme. Skymaps for $\gamma$-ray and $\nu$ signals from generic annihilation/decay spectra
are now direct outputs of \clumpy{}. Making use of \healpix{} routines, smoothing by a user-defined instrumental Gaussian beam and computing the angular power spectrum of the maps is now possible. In addition to these improvements, the main novelty is the implementation of a Jeans analysis
module, to obtain dark matter density profiles from kinematic data in relaxed spherical systems (e.g., dwarf
spheroidal galaxies). The code is also interfaced with the \great{}
toolkit designed for Markov Chain
Monte Carlo analyses, from which probability density functions and credible intervals can be obtained for velocity
dispersions, dark matter profiles, and $J$-factors.
\end{abstract}

\end{frontmatter}
%
% Computer program descriptions should contain the following
% PROGRAM SUMMARY.
%
{\bf PROGRAM SUMMARY}
  %Delete as appropriate.

\begin{small}
\noindent
{\em Program Title:} \clumpy{}                                   \\
%{\em Journal Reference:}                                      \\
  %Leave blank, supplied by Elsevier.
%{\em Catalogue identifier:}                                   \\
  %Leave blank, supplied by Elsevier.
%{\em Licensing provisions:} none                               \\
  %enter "none" if CPC non-profit use license is sufficient.
{\em Programming language:} C/C++                           \\
{\em Computer:} PC and Mac                                         \\
  %Computer(s) for which program has been designed.
{\em Operating system:} UNIX(Linux), MacOS X                    \\
  %Operating system(s) for which program has been designed.
{\em RAM:} between 500MB and 1GB depending on the size of the requested skymap        \\
  %RAM in bytes required to execute program with typical data.
  % Please give some freely chosen keywords that we can use in a
  % cumulative keyword index.
{\em Classification:} 1.1, 1.7, 1.9               \\
  %Classify using CPC Program Library Subject Index, see (
  % http://cpc.cs.qub.ac.uk/subjectIndex/SUBJECT_index.html)
  %e.g. 4.4 Feynman diagrams, 5 Computer Algebra.
{\em External routines/libraries:} CERN \rootcern{} (\url{http://root.cern.ch}), 
\gsl{} (\url{http://www.gnu.org/software/gsl}), 
\cfitsio{} (\url{http://heasarc.gsfc.nasa.gov/fitsio/fitsio.html}),
\healpix{} C++ and F90 (\url{http://healpix.sourceforge.net/index.php}),
\great{} (\url{http://lpsc.in2p3.fr/great}) (for MCMC analyses only), 
and \doxygen{} (\url{http://www.doxygen.org}) (optional) \\
  % Fill in if necessary, otherwise leave out.
{\em Nature of problem:} Calculation of dark matter profile from kinematic data,
$\gamma$-ray and $\nu$ signals from dark matter annihilation/decay.
\\
  %Describe the nature of the problem here.\\
{\em Solution method:} Solve the integro-differential Jeans equation (optimised for
speed) for several generic distributions (dark matter profile, light profile, velocity
anisotropy). Integration of the DM density (squared) along a line of sight for
generic dark matter haloes with substructures (spatial, mass, concentration
distributions).  Draw full skymaps of $\gamma$-ray and $\nu$ emission from dark matter
structures, smoothed by an instrument PSF using \healpix{} tools.
  %Describe the method solution here.
   \\
{\em Restrictions:}
The diffuse extragalactic contribution to the signal (and $\gamma$-ray attenuation)
as well as secondary radiation from dark matter remain to be included in order
to provide a comprehensive description of the expected signal.
  %Describe any restrictions on the complexity of the problem here.
   \\
{\em Running time:} This is highly dependent of the user-defined choices of DM profiles, precision $\epsilon$ and integration angle $\alpha_{\rm int}$:
\begin{itemize}
\item $\sim 1$\;hour for a full skymap (including substructures) with $\alpha_{\rm int}=0.1^\circ$ and $\epsilon=0.01$; 
\item $\lesssim 1$\;mn for a $5^\circ \times 5^\circ$ skymap (including substructures) with $\alpha_{\rm int}=0.1^\circ$ and $\epsilon=0.01$; 
\item $\sim 5$\;mn for a typical Jeans/MCMC analysis (on a `ultrafaint'-like dwarf spheroidal galaxy) using a constant
  anisotropy profile.
\end{itemize}

  %Give an indication of the typical running time here.
%\begin{thebibliography}{0}
%\bibitem{1}Reference 1         % This list should only contain those items referenced in the                 
%\bibitem{2}Reference 2         % Program Summary section.   
%\bibitem{3}Reference 3         % Type references in text as [1], [2], etc.
                               % This list is different from the bibliography at the end of 
                               % the Long Write-Up.
%\end{thebibliography}
%* Items marked with an asterisk are only required for new versions
%of programs previously published in the CPC Program Library.\\
\end{small}
\newpage
\setcounter{tocdepth}{2}
%\tableofcontents

%\linenumbers

%%%%%%%%%%%%%%%%%%%%%%%%%%%%%%%%%%%%%%%%%%%%%%%%%%%%%%%%%%%%%%%%%%%%%%%%%%%%%%
%%%%%%%%%%%%%%%%%%%%%%%%%%%%%%%%%%%%%%%%%%%%%%%%%%%%%%%%%%%%%%%%%%%%%%%%%%%%%%%% main text
\section{Introduction \label{sec:intro}}

Dark matter indirect detection aims at measuring the end products of dark matter (DM)
annihilation or decay ($e^+$, $\bar{p}$, $\bar{d}$, $\gamma$, $\nu$). The recent
results from the PAMELA \cite{2009Natur.458..607A,2010PhRvL.105l1101A} and AMS-02
\cite{2013PhRvL.110n1102A} experiments for charged particles, and the wealth of Fermi-LAT results for $\gamma$-rays
\cite{2012ApJ...761...91A,2013PhRvD..88h2002A,2014PhRvD..89d2001A}, show that we are
starting to probe the region of interest of the parameter space for new physics.
Due to the complexity of the signal and backgrounds involved, the need for public
tools for cross-checks and progresses in the field is mandatory.

Several particle physics public tools exist to calculate the spectrum
of species produced from dark matter annihilation and decay (e.g., {\tt micrOMEGAs}
\cite{2011CoPhC.182..842B,2014CoPhC.185..960B}, {\tt DarkSUSY}
\cite{2004JCAP...07..008G}). The astrophysical side of the calculation depends on the nature of the
created particle: for charged species ($e^+$, $\bar{p}$, $\bar{d}$), a
diffusion/convection equation must be solved, and several propagation codes are
available (e.g., {\tt GalProp}\footnote{\url{http://galprop.stanford.edu}} and {\tt
Dragon}\footnote{\tt \url{http://www.dragonproject.org}}). Neutral particles ($\gamma$,
$\nu$) propagate in straight lines, and the main uncertainty in the signal is related
to the $J$-factor calculation from DM structures and substructures: \clumpy{}\footnote{\tt \url{http://lpsc.in2p3.fr/clumpy}} \cite{2012CoPhC.183..656C}
(Paper~I) is the only public code for the full $J$-factor calculation.

\clumpy{} has been used to calculate $J$-factor in DM haloes of dwarf
spheroidal galaxies (dSphs)
\cite{2011ApJ...733L..46W,2011MNRAS.418.1526C}, DM-supported
H{\sc i} clouds \cite{2014MNRAS.442.2883N}, and towards the Galactic
centre to place limits on DM annihilation using the ANTARES neutrino telescope \cite{2015arXiv150504866A}. The code has been 
extended to provide additional outputs (sorting, {\em population} study) when
applied on a sizeable sample of galaxy cluster haloes
\cite{2012PhRvD..85f3517C,2012MNRAS.425..477N,2012A&A...547A..16M}, or galactic subhaloes.  The second 
\clumpy{} release, presented in this paper, includes these additional outputs, but also provides a
better description of several quantities related to DM haloes and their substructures,
as well as more flexible and useful outputs. The main
novelties of the code are the following:
\begin{enumerate}
  \item implementation of the well-established Jeans analysis \cite{2008gady.book.....B}
  to extract DM profiles from the kinematics of baryonic tracers. It is interfaced
  with a Markov Chain Monte Carlo (MCMC) engine which enables the user
  to perform the complete analysis from kinematic data to the
  astrophysical factors probability density distributions;

  \item triaxiality of the DM haloes (Milky Way, or any other DM halo)
    can now be enabled;

  \item use of the \healpix{} ({\sc H}ierarchical {\sc E}qual
  {\sc A}rea iso{\sc L}atitude {\sc Pix}elation)
  and \cfitsio{} input/output libraries
  for 2D skymaps \fits{} outputs, including smoothing by a Gaussian beam and power spectrum tools
  (the code for 2D skymaps has also been optimised for speed);
  
  \item use of tabulated DM $\gamma$-ray and $\nu$ spectra
  from \citep{2011JCAP...03..051C}\footnote{\tt \url{http://www.marcocirelli.net/PPPC4DMID.html}}
  to compute fluxes for annihilation and decay ($\nu$ oscillations are included).
\end{enumerate}

This paper highlights \clumpy{}'s main features, old and new; a more thorough
description (and examples) can be found in the fully \doxygen{}-documented code. The
paper is organised as follows. Section~\ref{sec:calculation} briefly recalls the
formalism for calculating the $J$-factor in the presence of dark matter substructures.
Section~\ref{sec:updatesJ} provides updated formulae for this new version, which includes
a statistical description of the concentration parameter, the multi-level boost
calculation, and triaxiality for DM haloes. Section~\ref{sec:outputs} focuses on
the description of the new outputs provided in this version (\fits{} files containing maps using the
\healpix{} pixelisation scheme, $\gamma$-ray and $\nu$ fluxes). Section~\ref{sec:jeans} deals
with the Jeans analysis, its implementation in \clumpy{}, and the interface with the MCMC engine \great{}.
Run examples with a short description of \clumpy{} (and its installation) are given in Section~\ref{sec:examples}. We conclude in Section~\ref{sec:conclusion}.

%%%%%%%%%%%%%%%%%%%%%%%%%%%%%%%%%%%%%%%%%%%%%%%%%%%%%%%%%%%%%%%%%%%%%%%%%%%%%%
%%%%%%%%%%%%%%%%%%%%%%%%%%%%%%%%%%%%%%%%%%%%%%%%%%%%%%%%%%%%%%%%%%%%%%%%%%%%%%
\section{Dark matter annihilation or decay: reminder\label{sec:calculation}}

          %%%%%%%%%%%%%%%%%%%%%%%%%%%%%%%%%%
\subsection{Fluxes}

The $\gamma$-ray or $\nu$ flux $\ud \Phi_{\gamma,\nu}/\ud E$ from dark matter annihilating/decaying
particles is expressed as the product of a particle physics term by an astrophysical
contribution $J$. At energy $E$ and in the direction $(\psi,\theta)$, the flux
integrated over the solid angle  $\Delta \Omega=2\pi\,(1-\cos\,\alpha_{\rm int})$ is
given by
\beq
\frac{\ud \Phi_{\gamma,\nu}}{\ud E}(E,\psi,\theta,\Delta
\Omega)=\frac{\ud \Phi^{PP}_{\gamma,\nu}}{\ud E}(E)\times J(\psi,\theta,\Delta \Omega) \,,
\label{eq:flux-general}
\eeq
in which $\ud\Omega=\ud\beta \sin\alpha \ud\alpha$ is the elementary solid angle around
the line-of-sight direction $\psi,\theta$ (longitude and latitude in
Galactic coordinates)\footnote{A sketch of the coordinate framework is provided in the code
documentation (see also Fig.~6 of Paper~I).}.

\subsubsection{Particle physics term} The particle physics term depends on whether the DM candidate annihilates or decays. In this version
(as in the previous one), we only consider the continuum emission \cite[e.g.,][]{2011JCAP...03..051C}:
\begin{equation}\label{eq:term-pp}
  \frac{\ud \Phi_{\gamma,\nu}}{\ud E}(E)
      = \frac{1}{4\pi}\,\sum_{f}\frac{dN^{f}_{\gamma,\nu}}{dE}\, B_{f}
        \times
    \begin{cases}
       \displaystyle \frac{\langle \sigma_{\rm ann}v \rangle}{m_{\rm DM}^{2}\delta}& \text{(annihilation)}\\ \\
       \displaystyle \frac{1}{\tau_{\rm DM}\;m_{\rm DM}} & \text{(decay)}
    \end{cases} 
\end{equation}
with $m_{\rm DM}$ the mass of the DM candidate, $B_f$ the branching ratio into the final state $f$ and its
yield per reaction $dN^{f}_{\gamma,\nu}/dE$ (see \S\ref{sec:gammaray_nu}), and
    \begin{itemize}
      \item $\sigma_{\rm ann}$ is the annihilation cross section, and $\langle\sigma_{\rm ann}v\rangle$
        the annihilation rate averaged over the DM velocity distribution, $\delta$ equals 2 (resp. 4) for a Majorana
        (resp. Dirac) fermion;
      \item $\tau_{\rm DM}$ is the decay lifetime.
    \end{itemize} 

\subsubsection{Astrophysical $J$ (annihilation) or $D$ (decay) factor}
The astrophysical factor relies on the
integration over the solid angle $\Delta \Omega$ of some power of the DM density 
$\rho(\psi,\theta, l,\alpha,\beta)$ at coordinate ($l,\alpha,\beta)$ in the 
line-of-sight direction $(\psi, \theta)$:
\begin{equation}\label{eq:term-astro}
J(\psi,\theta, \Delta \Omega) = \int_{0}^{\Delta \Omega}\int_{\rm{l.o.s}} \!\!\ud l \, \ud \Omega \times
    \begin{cases}
       \displaystyle \,\rho^2& \text{(annihilation)}\\
       \displaystyle \,\rho  & \text{(decay)}\;.
    \end{cases} 
\end{equation}
Note that, depending on the community, $J$- and $D$-factors are either expressed in {\em astrophysics} units
($M_\odot^2\;{\rm kpc}^{-5}$ and $M_\odot\;{\rm kpc}^{-2}$ respectively) or {\em particle physics} units
(GeV$^2$\;cm$^{-5}$ and GeV\;cm$^{-2}$ respectively). All calculations
in \clumpy{} are performed in {\em astrophysics} units,
but a new keyword introduced in this version ({\tt gSIMU\_IS\_ASTRO\_OR\_PP\_UNITS}) allows
the user to select the preferred units for the outputs (plots, {\tt
  ASCII} and \fits{} files).

In the following, we concentrate on the $J$-factor calculation for annihilation, for
which the contribution from substructures is able to boost the signal (there is no
boost for decaying DM). In both cases, the formalism is the same and is implemented
similarly in \clumpy{}.

          %%%%%%%%%%%%%%%%%%%%%%%%%%%%%%%%%%
\subsection{$J$-factor and substructures: formalism}
Here, we briefly
recap the formalism used to take into account substructures in
\clumpy{}. We refer the reader to Paper~I for a more detailed description.
The changes brought by this new
release regarding the handling of substructures are postponed to \S\ref{sec:updatesJ}.
%Since \clumpy{} can either rely on the mean value of the signal from all the subhaloes or 
%calculate explicitly the contributing of the meaningful ones, we come back to this subtlety
%in this section.

%%%%%%%%%%%
\subsubsection{Formal description}
From Eq.~(\ref{eq:term-astro}), the total DM density $\rho$ must be known at each
position. In the $\Lambda$CDM cosmological model, structures form
in a bottom-up manner: micro-haloes form first, larger ones collapse later, and this
process, accompanied with merger events, lead to the global picture of clumps within clumps
within clumps, etc. Each DM clump can be seen as a density peak inside its host halo,
and it is therefore convenient to separate, for a given halo, the main
distribution (called smooth halo) from the contributions of each clump.

The astrophysical contribution to the annihilation flux is thus explicitly written to
be
\beq
J = \int_0^{\Delta\Omega}\int_{l_{\rm min}}^{l_{\rm max}} \left( \rho_{\rm sm} + \sum_i \rho_{\rm
    cl}^i\right)^2 \ud l \, \ud\Omega\;,
    \label{eq:tot_J}
\eeq
where $\rho_{\rm cl}^i$ corresponds to the inner density of the $i$-th clump contained in the volume element. Three terms arise from this equation
(smooth only, clumps contribution, and cross-product):
\begin{eqnarray}
   J_{\rm sm} &\equiv& \int_0^{\Delta\Omega}\int_{l_{\rm min}}^{l_{\rm max}} 
      \rho_{\rm sm}^2 \ud l \, \ud\Omega\;, 
       \label{eq:gal_Jsm}
  \\
   J_{\rm subs} &\equiv& \int_0^{\Delta\Omega}\int_{l_{\rm min}}^{l_{\rm max}}
       \left(\sum_i \rho_{\rm cl}^i\right)^2 \ud l \, \ud\Omega\;,
       \label{eq:gal_Jsub}
  \\
   J_{\rm cross-prod} &\equiv& 2\int_0^{\Delta\Omega}\int_{l_{\rm min}}^{l_{\rm max}}
       \rho_{\rm sm}\sum_i \rho_{\rm cl}^i \ud l \, \ud\Omega\;.
       \label{eq:gal_Jcrossprod}
\end{eqnarray}

The calculation of $J_{\rm subs}$ and $J_{\rm cross-prod}$ 
described in Paper~I was using only one level of substructures. This
new release of the code allows the inclusion of more levels of
substructures as described in \S\ref{subsec:m_c}.

%%%%%%%%%%%
\subsubsection{Continuum limit}

A typical DM halo of $10^{12}M_\odot$ (Milky-Way like) contains up to $10^{14}$
substructures, which renders the explicit calculation of the above sums prohibitive.
This huge number allows the use of the continuum limit as the clump
positions and masses are random variables, drawn from distribution
functions obtained by N-body numerical simulations and/or semi-analytical
calculations\footnote{Large uncertainties remain on these distributions, all the more
because small halo masses are not resolved, even in the most
computationally heavy simulations: \clumpy{} is partly designed to enable quick
calculations of the $J$-factor for any input distribution, in order to check the
sensitivity/robustness of the results against the uncertain parameters of the
distributions.}. 

As detailed in \S\ref{subsec:m_c}, the above equations can often be replaced by
averaged ones. In particular, the total averaged density corresponds to
\beq
\langle\rho_{\rm tot}\rangle (r) = \rho_{\rm sm}(r) + \langle \rho_{\rm
  subs}\rangle(r),
\label{eq:rhotot}
\eeq
where $ \rho_{\rm sm}(r)$ is the smooth component and $\langle \rho_{\rm sub}\rangle(r)$ 
the spherical shell average density of substructures (at each radius $r$). 
The total averaged spherical shell density $\langle\rho_{\rm tot}\rangle (r)$
is usually parametrised by Zhao or Einasto profiles (see \S3.3.4 of Paper~I and {\tt profiles.h}).
A saturation density $\rho_{\rm sat}$ ({\tt gDM\_RHOSAT}) provides a cut-off radius
below which the annihilation rate is constant (equilibrium between free fall time
and annihilation time). For instance, selecting the total average density profile
$\langle\rho_{\rm tot}\rangle$ and the clump distribution parameters
$\langle \rho_{\rm subs}\rangle$ (see next paragraph), Eq.~(\ref{eq:rhotot})
is used to get the smooth distribution $\rho_{\rm sm}(r)$ that is plugged
in Eq.~(\ref{eq:gal_Jsm}).

%%%%%%%%%%%
\subsubsection{Validity of the mean description}

At first order, a random variable (e.g. the mass and position of
substructures) is described by its average value and variance. Departure
from the average can arise if a small number of objects contribute significantly to
the total $J$-factor, which happens if a massive clump dominates, or if one of the
smallest halo (the smaller, the more numerous they are) is sitting almost at the
observer location. The latter configuration only happens for clumps in the
Galaxy, since clumps in dSphs or extragalactic objects are far away.

As presented in Paper~I, for $J$-factor skymaps of the Galaxy, {\sc
\clumpy{}} relies on a combination of the calculation of the average signal 
and the calculation of individual drawn clumps
above and below a
critical distance $l_{\rm crit}$ (the computation of which is detailed in Paper~I\footnote{The critical distance is obtained by requiring the relative error of the signal integrated from $l_{\rm crit}$ to remain lower than a user-defined precision requirement.}),
respectively. This strategy
ensures a controlled and extremely quick calculation of skymaps: the number of clumps
to draw in the Galaxy is reduced from a few tens of thousands to a
few hundreds (see
table~1 of Paper~I) depending on the user-chosen level of precision (or more
precisely, the level of fluctuation selected w.r.t. the mean signal). For all other
objects|like dSphs, galaxies, galaxy
clusters|, the mean description is usually sufficient.

%%%%%%%%%%%%%%%%%%%%%%%%%%%%%%%%%%%%%%%%%%%%%%%%%%%%%%%%%%%%%%%%%%%%%%%%%%%%%%
%%%%%%%%%%%%%%%%%%%%%%%%%%%%%%%%%%%%%%%%%%%%%%%%%%%%%%%%%%%%%%%%%%%%%%%%%%%%%%
\section{Updates and novelties in the $J$-factor computation }
\label{sec:updatesJ}
Three major improvements have been included: the concentration of the clumps is now
dealt with a distribution function, so as to include an uncertainty around the
mass-concentration relationship, and the calculation
of the boost due to substructures can include, if required, several levels of
substructures (\S\ref{subsec:m_c}); it is now possible to
deal with triaxial DM haloes in addition to spherical ones (\S\ref{subsec:triaxiality}).

          %%%%%%%%%%%%%%%%%%%%%%%%%%%%%%%%%%
\subsection{Improved concentration description and subhalo levels}
\label{subsec:m_c}

For a given DM profile, the physical properties of a subhalo are fully defined by its position, mass, and
concentration\footnote{The concentration at a given characteristic overdensity $\Delta$
is defined to be $c_\Delta\equiv R_\Delta/r_{-2}$, where $R_\Delta$ is the radius of
the clump for which the density equals this overdensity, and $r_{-2}$ is the position
where the logarithmic slope of the DM density profile of the clump reaches -2.}. In the previous \clumpy{}
release, the position and mass were random variables of
user-defined distribution functions. The mass-concentration
relation was fixed so that two DM haloes of equal mass would
have had the same concentration. Numerical simulations have however shown significant
uncertainties in the determination of the mass-concentration relation,
which could be parametrised by a dispersion around an average
relation. Therefore, in this release, the
concentration parameter is a new random variable, characterised by a
specific distribution function as described below.

%%%%%%%%%%%
\subsubsection{Substructures: random variables and distributions}
For a total number of clumps $\Ntot$ in a host halo, the
substructure distribution is modelled by:
\beq
   \frac{\ud^3N}{\ud V \ud M\ud c} = \Ntot
     \frac{\ud{\cal P}_V}{\ud V}(r)\cdot\frac{\ud {\cal P}_M}{\ud M}(M) \cdot
      \frac{\ud{\cal P}_c}{\ud c}(M,c) \textrm{.}
\label{eq:distribution-clumps}
\eeq
In the above equation, each distribution $\ud {\cal P}$ is a probability, i.e. is
normalised to 1 when integrated on its domain of applicability. In terms of
parametrisation:
\begin{itemize}
  \item as in the previous release, the spatial distribution $\ud{\cal P}_V(r)/\ud V$ is 
        selected from {\tt gENUM\_PROFILE}  (see table~\ref{tab:enum});
  \item the mass distribution $\ud {\cal P}_M(M)/\ud M$ is a simple power-law with two
        parameters (normalisation and slope $\alpha_M\approx 1.9$),
        again similarly to the previous release;
  \item the concentration distribution $\ud{\cal P}_c(M,c)/\ud c$ is a
    new feature of the code (in {\tt clumps.h}) and is chosen to be either
    a Dirac function or a log-normal distribution ({\tt gENUM\_CVIR\_DIST},
    see table~\ref{tab:enum}):
   \begin{equation}
     \frac{\ud {\cal P}_c}{\ud c}(M,c) = \frac{\exp^{\displaystyle 
             -\left [\frac{\ln{c} - 
              \ln({\bar{c}(M)})}{\sqrt{2}\sigma_c(M)} \right ]^2}}{\sqrt{2 \pi}\,\,c\,\,\sigma_c(M)}.
   \label{eq:dpdc}  
   \end{equation}
   The latter case means that the concentration of a clump of mass $M$ is randomly drawn
   from the above distribution around the mean concentration
   $\bar{c}(M)$ (calculated from a {\tt gENUM\_CVIRMVIR} enumerator), with a scatter $\sigma_c(M)$ (e.g., $0.14 - 0.18$ \cite{2001MNRAS.321..559B,2002ApJ...568...52W,2014MNRAS.442.2271S}). 
   The parameters in {\tt clumpy\_params.txt} (see table~\ref{tab:param})
   are {\tt gDM\_FLAG\_CVIR\_DIST} and {\tt gDM\_LOGCVIR\_STDDEV}.
\end{itemize}

%%%%%%%%%%%
\subsubsection{\clumpy{} formulae to calculate $\langle J\rangle$}
The description of the concentration in terms of a distribution
function implies some modifications to the average $\langle J\rangle$
as defined in paper~I (see \S3.3.6 and the documentation in {\tt clumps.h} 
for more details). The formulae below recap and extend Eqs.~(17)~--~(22) of Paper~I. The
average mass and the mean of (some power of) the distance are left
unchanged and read
\beq
\!\!\langle M\rangle \!=\!\!\! \int_{M_{\rm min}}^{M_{\rm max}}\!\!\!\!\! 
  M\frac{\ud {\cal P}_M}{\ud M} {\ud M},
\;\;\;
\langle l^n \rangle \!=\!\!\! \int_0^{\Delta\Omega}\!\!\!\! \int_{l_{\rm min}}^{l_{\rm
    max}}\!\!\!\! l^{\,n} \frac{\ud {\cal P}_V}{\ud V} l^{\,2} \ud l \, \ud\Omega.
%\label{eq:mean_dist}
\eeq
The intrinsic luminosity of a clump ${\cal L}(M,c)$ now depends on the
concentration $c$,
\beq
\!\!{\cal L}(M,c)\!\equiv\!\!\int_{\rm V_{\rm cl}}\!\! \rho_{\rm cl}^2(M,c) \,\ud V\,,
\label{eq:luminosity}
\eeq
so that the mean luminosity (over a range of $M$ and $c$) becomes
\beq
\!\!\langle {\cal L} \rangle \!=\!\!\!
     \int_{M_{\rm min}}^{M_{\rm max}} \!\!\frac{\ud {\cal P}_M}{\ud M}(M)
     \!\!\!
     \int_{c_{\rm min}(M)}^{c_{\rm max}(M)} \!\!\frac{\ud {\cal P}_c}{\ud c}(M,c) 
     \,{\cal L}(M,c)\, \ud c \,\ud M,
     \label{eq:L_with_c}
\eeq
and the average $J$ value from the clumps becomes
\begin{eqnarray}
 \!\! \langle J_{\rm subs}\rangle \!\!\!&=&\!\!\!\! \Ntot\!\! 
 \int_0^{\Delta\Omega} \!\!
   \int_{l_{\rm min}}^{l_{\rm max}}
       \frac{{\rm d}{\cal P}_V}{\ud V}(l,\Omega) \,\ud l \,\ud\Omega 
    \int_{M_{\rm min}}^{M_{\rm max}}\! \frac{\ud {\cal P}_M}{\ud M}(M)
    \nonumber\\
    \!\!\!&\times&\!\!\!\!
    \int_{c_{\rm min}(M)}^{c_{\rm max}(M)} \!\!\frac{\ud {\cal P}_c}{\ud c}(M,c) 
     \,{\cal L}(M,c)\, \ud c \,\ud M.
\label{eq:gal_meanJcl} 
\end{eqnarray}
The variance $\sigma^2_{\rm subs}$ on $J$ remains as calculated in Paper~I, i.e.
\beq
\sigma^2_{\rm subs}= \langle {\cal L}^2\rangle \left\langle\frac{1}{l^4}\right\rangle 
 - \langle{\cal L}\rangle^2 \left<\frac{1}{ l^2}\right>^2,
\label{eq:var_J}
\eeq
but the difference is that $\langle{\cal L}^2\rangle$ and $\langle{\cal L}\rangle$
must now be calculated accounting for the extra integration on the distribution
of concentration  $\ud{\cal P}_c(M,c)/\ud c$. Using {\tt gENUM\_CVIR\_DIST=kLOGNORM}
with a typical scatter of $0.18$ around $\langle c_{\rm vir}\rangle$~\cite{2001MNRAS.321..559B}
leads to an increase of $\langle{\cal L}\rangle$ by $\sim 15\%$ at all masses
(see the documentation in {\tt clumps.h} for illustrative plots), compared
to the use of {\tt gENUM\_CVIR\_DIST=kDIRAC} (as in the previous release).

          %%%%%%%%%%%%%%%%%%%%%%%%%%%%%%%%%%
\subsubsection{Substructure boost: including extra-levels}
\label{subsubsec:extralevels}
Based on the hierarchical structure formation scenario, the
calculation of $J_{\rm subs}$ and $J_{\rm cross-prod}$ should include
a multi-level description of the clumps, i.e. contributions from 
clumps within clumps within clumps, etc.

To exemplify how this is implemented in \clumpy{}, let us consider a
DM halo, with density profile $\rho^{\rm tot}_{\rm cl}$, fully encompassed in the integration angle. 
Not considering any substructure within this halo, its intrinsic
luminosity is directly given by Eq.~(\ref{eq:luminosity}), denoted
`0' to highlight the absence of substructures in that DM halo:
\beq
{\cal L}_{0}(M,c)\equiv\int_{\rm V_{\rm cl}} \left[\rho_{\rm cl}^{\rm tot}(M,c)\right]^2 \,\ud V\,.
\eeq
A hierarchy of $n$ levels of substructures within this host halo
($n=1,2,3$, etc.), may be computed from the $n-1$ level as (we drop the implicit dependence on $c$ for compactness):
\begin{eqnarray}
\label{eq:multi-level}
{\cal L}_{n}(M) &=& {\cal L}_{\rm sm}(M) + {\cal L}_{\rm cross-prod}(M)\\
    &+& N_{\rm tot}(M)\int_{M_{\rm min}}^{M_{\rm max}(M)} \!\!\!\!{\cal L}_{n-1}(M')\frac{\ud{\cal P}}{\ud M'}(M') dM'
    \,.\nonumber
\end{eqnarray}
with 
\begin{eqnarray}
   {\cal L}_{\rm sm}(M)&\equiv&\int_{\rm V_{\rm cl}} \left[\rho_{\rm cl}^{\rm sm}(M)\right]^2 \,\ud V\,;\nonumber\\
   {\cal L}_{\rm cross-prod}(M)&\equiv&2\int_{\rm V_{\rm cl}} \rho_{\rm cl}^{\rm sm}(M)\,\langle\rho_{\rm subs}(M)\rangle \,\ud V\,.\nonumber
\end{eqnarray}
In the first version of the code, only $n=1$ was considered. This has
now been extended to any level with a recursive implementation of Eq.~(\ref{eq:multi-level}). Part of this implementation requires to identify the radius at which the slope of the spatial distribution of the clumps equals $-2$, i.e., it can only be enabled with distributions whose outer slope is steeper than $-2$.

\begin{figure}[!t]
\centering
\includegraphics[width=\columnwidth]{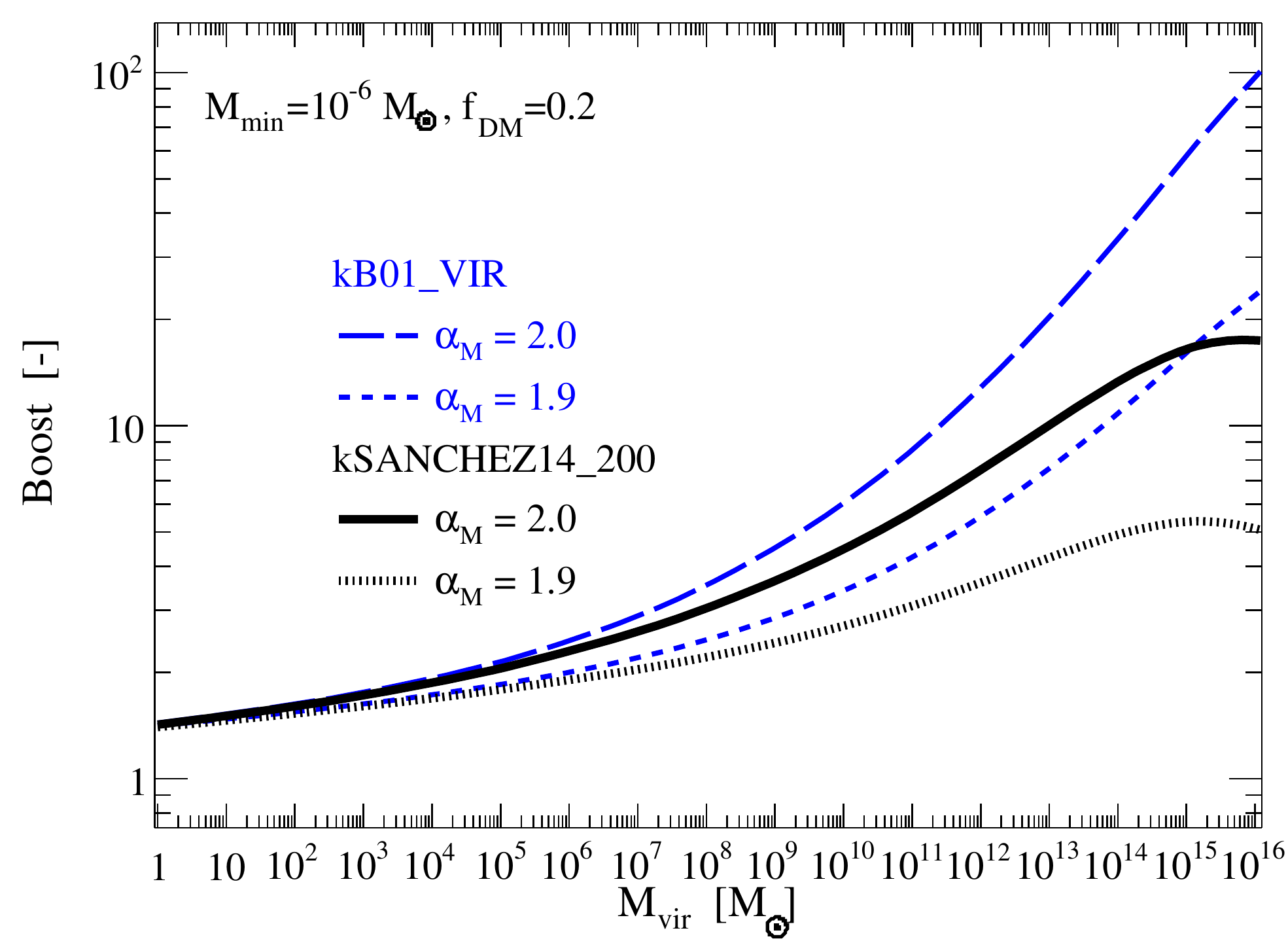}
\includegraphics[width=\columnwidth]{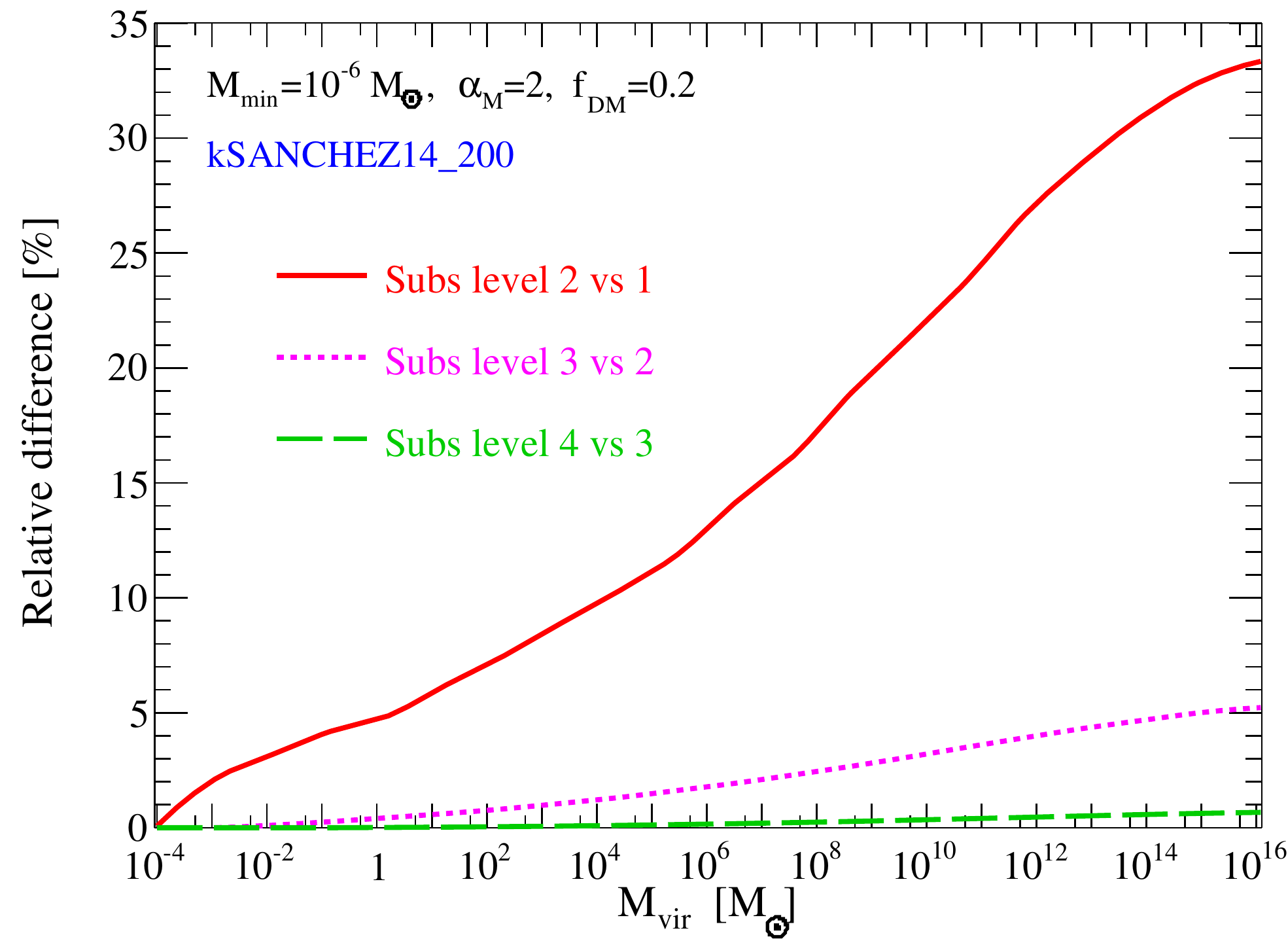}
\caption{{\em Top panel:} boost factor ${\cal L}_4/{\cal L}_0$ as a function of the host halo mass for two $\alpha_M$
and two $c_{\rm vir}-R_{\rm vir}$ relationships. {\em Bottom panel:} relative importance of
higher-level substructure contributions.}
\label{fig:boost_multi}
\end{figure}

The impact of the substructures as well as the relative contributions of higher level
subhaloes are illustrated in Fig.~\ref{fig:boost_multi}. The top panel corresponds to
the boost factor ${\cal L}_4/{\cal L}_0$ calculated from Eq.~(\ref{eq:multi-level})
as a function of the host halo
mass $M_{\rm vir}$. The exact dependence depends on several key parameters (minimal
mass of subhaloes, mass distribution slope $\alpha_M$, and
$c_{\rm vir}-R_{\rm vir}$ parametrisation), but our results using, e.g., 
{\tt kSANCHEZ14\_200} for $c_{\rm vir}-R_{\rm vir}$ are in agreement with the results
of \cite{2014MNRAS.442.2271S}. The bottom panel details the contributions of higher-order
contributions with respect to the previous level. The main contribution
(not shown) is the first level of subhaloes (i.e. $n=1$), which is
responsible for most of the boost seen in the top panel. The second level of substructures further contributes
at most to 30\% (solid red line), for the most massive haloes. As
underlined in several previous studies using a different
approach \cite{2009JCAP...06..014M,2014MNRAS.442.2271S}, the third level of substructures
contributes to less than 5\% of the total.

\subsubsection{Recommendation regarding concentration and number of substructure levels}
\paragraph{Concentration parametrisations} The {\tt gDM\_FLAG\_CVIR\_DIST} parameter allows the user to choose the mass-concentration from eight pre-defined parametrisations taken from various studies \citep{2001MNRAS.321..559B,2014MNRAS.442.2271S,2001ApJ...554..114E,2007MNRAS.381.1450N,2008MNRAS.390L..64D,2010A&A...524A..68E,2012MNRAS.423.3018P,2012MNRAS.422..185G} and listed in Table~\ref{tab:enum}. While any of these parametrisations may be used to characterised the subhaloes concentration in \clumpy{}, it is important to keep in mind that, to date, these parametrisations have been established for main haloes only (i.e. not substructures). 
Furthermore, note that some of these parametrisations are simple power-laws, shown to provide good fits to simulations in the mass range $\sim 10^{10} - 10^{15} M_\odot$ \citep{2007MNRAS.381.1450N,2008MNRAS.390L..64D} or to X-ray galaxy cluster data \citep{2010A&A...524A..68E}; these should however not be extrapolated to lower masses where they would overestimate the concentration. Parametrisations allowing for a flattening of the mass-concentration relation at lower mass such as \cite{2001MNRAS.321..559B,2014MNRAS.442.2271S,2001ApJ...554..114E,2012MNRAS.422..185G} are more realistic and the default (and recommended) \clumpy{} setting is {\tt gDM\_FLAG\_CVIR\_DIST=kSANCHEZ14\_200} \citep{2014MNRAS.442.2271S}.

\paragraph{Number of substructure levels} The level parameter ({\tt gDM\_SUBS\_NUMBEROFLEVELS} in {\tt
  clumpy\_params.txt}) is set to $n=1$ by default. Setting
$n=0$ is not implemented directly but is obtained instead by asking
that the fraction of the mass of the host halo under the form of
substructure is zero ({\tt gTYPE\_SUBS\_MASSFRACTION = 0} in {\tt
  clumpy\_params.txt}). 

Note that enabling more than one
level of substructures ({\tt gDM\_SUBS\_NUMBEROFLEVELS} in {\tt
  clumpy\_params.txt}), and/or requiring a log-normal distribution of concentrations
({\tt gENUM\_CVIR\_DIST=kLOGNORM}), significantly increases the computational time
required to run \clumpy{}. We find that using the concentration scatter increases by $\sim 30\%$ the computational time for a given skymap. Asking for a second level of substructures almost doubles the duration of the run. Requiring simultaneously the concentration scatter and a second level of substructures increases by a factor $\sim 10$ the computational time. We underline that using $n=2$ multi-levels
is enough to reach a precision better than 5\%.

          %%%%%%%%%%%%%%%%%%%%%%%%%%%%%%%%%%
\subsection{Host halo triaxiality}
\label{subsec:triaxiality}
\begin{table*}
\caption{Enumerators and allowed keyword (and reference) in the \clumpy{} code. Highlighted entries (shade of grey) correspond to
new enumerators and/or keywords of this version (w.r.t. the first release).}
\begin{center}
{
\small
%\footnotesize
\begin{tabular}{ll}  \hline
Enumerator & Flags available\\ \hline\rowcolor{lightgray}
{\tt gENUM\_ANISOTROPYPROFILE}\!\!\!  & {\tt kCONSTANT\!, kBAES\,\cite{2007A&A...471..419B}\!, kOSIPKOV\,\cite{1979PAZh....5...77O,1985AJ.....90.1027M}}\\\rowcolor{lightgray}
{\tt gENUM\_LIGHTPROFILE}       & {\tt kEXP2D\,\cite{2009MNRAS.393L..50E}\!, kEXP3D\,\cite{2009MNRAS.393L..50E}\!, 
                                  kKING2D\,\cite{1962AJ.....67..471K}\!, kPLUMMER2D\,\cite{1911MNRAS..71..460P}\!,
                                  kSERSIC2D\,\cite{1968adga.book.....S}\!, kZHAO3D\,\cite{1990ApJ...356..359H,1996MNRAS.278..488Z}}\!\!\!\!\\
%
%&\\[-0.12cm]
{\tt gENUM\_CVIRMVIR}           & {\tt kB01\_VIR\,\cite{2001MNRAS.321..559B}\!, kENS01\_VIR\,\cite{2001ApJ...554..114E}\!,
                                  kNETO07\_200\,\cite{2007MNRAS.381.1450N}\!, 
                                  kDUFFY08F\_\{VIR,\,200,\,MEAN\}\,\cite{2008MNRAS.390L..64D}}\\
                                & {\tt kETTORI10\_200\,\cite{2010A&A...524A..68E}\!,\,kPRADA11\_200\,\cite{2012MNRAS.423.3018P}\!,\,
                                  kGIOCOLI12\_VIR\,\cite{2012MNRAS.422..185G}\!,\,kSANCHEZ14\_200\,\cite{2014MNRAS.442.2271S}}\\\rowcolor{lightgray}
{\tt gENUM\_CVIR\_DIST}         & {\tt kLOGNORM\,\cite{2001MNRAS.321..559B}\!,\,\tt kDIRAC}\\
%
%&\\[-0.2cm]
{\tt gENUM\_PROFILE}            & {\tt kZHAO\,\cite{1990ApJ...356..359H,1996MNRAS.278..488Z}\!,\,kEINASTO\,\cite{2004MNRAS.349.1039N}\!,\,kEINASTO\_N\,\cite{2006AJ....132.2685M}\!,
                                   \!\!\colorbox{lightgray}{kBURKERT\,\cite{1995ApJ...447L..25B}\!,}}\\[-0.05cm]
                                & {\tt \colorbox{lightgray}{kEINASTOANTIBIASED\_SUB\,\cite{2007ApJ...671.1135K,2015MNRAS.447..939L}\!,\,kGAO\_SUB\,\cite{2004MNRAS.355..819G,2008ApJ...679.1260M}}}\\
{\tt gENUM\_TYPEHALOES}         & {\tt kDSPH, kGALAXY, kCLUSTER}\\
{\tt gENUM\_FINALSTATE}         & {\tt kGAMMA, kNEUTRINO}\\\rowcolor{lightgray}
{\tt gENUM\_NUFLAVOUR}          & {\tt kNUE, kNUMU, kNUTAU}\\\rowcolor{lightgray}
{\tt gENUM\_PP\_SPECTRUMMODEL}\!\!\!\!\!\!  & \!\!\!{\tt (kBERGSTROM98\,\cite{1998APh.....9..137B}\!,\,kTASITSIOMI02\,\cite{2002PhRvD..66h3006T}\!,\,kBRINGMANN08\,\cite{2008JHEP...01..049B}\!)$^\star$,\,kCIRELLI11\_\{EW,\,NOEW\}\,\cite{2011JCAP...03..051C}}\\
\hline\hline
\end{tabular}
}
\label{tab:enum}
{\\\footnotesize $^\star$ Spectra for $\gamma$-rays only.}
\end{center}
\vspace{-0.3cm}
\end{table*}

Both numerical simulations and observations hint at triaxial rather
than spherical DM haloes \cite{2002ApJ...574..538J}. In this \clumpy{} release, we supplement the default spherical halo
configuration with a more general triaxial model $(a\neq b\neq c)$,
where $a$, $b$ and $c$ correspond to dimensionless major, intermediate and minor
axes. Cosmological simulations have shown that these
axes can vary as a function of the iso-density radius $R_{\rm iso}$ \citep{2002ApJ...574..538J},
but this refinement is not considered here. The position $(X,Y,Z)$ in the coordinate framework
attached to a given DM halo centre corresponds to the iso-density radius
\begin{equation}
 R_{\rm iso} =\sqrt{\frac{X^2}{a^2} + \frac{Y^2}{b^2} + \frac{Z^2}{c^2}}.
 \label{eq:Riso}
\end{equation}
In this configuration, the DM density is given by $\rho(R)$, where
$\rho$ can be any of the spherical profiles already implemented in \clumpy{} (e.g., Einasto, Zhao, \dots).

\subsubsection{Different implementations for different DM halo types}
The treatment of triaxiality relies on seven parameters (three axes for the shape, three Euler rotation angles,
for the orientation, and one boolean to switch on or off
triaxiality). 
\begin{itemize}
   \item Galaxy: for the Milky Way, these parameters are denoted {\tt gGAL\_TRIAXIAL\_XXX} and
   must be defined in the parameter file {\tt clumpy\_params.txt} (see table~\ref{tab:param}).
   Whenever triaxiality in enabled, the options propagates to the total, smooth and average clump contributions.
   \item Halo list: \clumpy{} can also run on a user-defined list of haloes. In addition to
   the position and structural parameters of each halo, already
   required in the first version of the code, the halo list may now include (as an
   option) the seven extra parameters for triaxiality (compare 
   {\tt data/list\_generic.txt} and {\tt data/list\_generic\_triaxial.txt}). Each halo in
   the list can therefore have its own triaxial properties.
   See the documentation for the required format and examples.
   \item Substructures: triaxiality is not enabled for the description of subhaloes
     in its host halo, since it would require a statistical distribution
   of axis ratios and orientations. First, the orientation is not important if the integration angle
   encompasses the subhalo volume (and this is the case for all subhaloes but a tiny fraction). Second,
   the exact distribution of axis ratios is not known, but the impact in the calculation is certainly
   sub-dominant compared with other uncertainties (concentration, profile, etc.).
\end{itemize}

\subsubsection{New \clumpy{} functions}
From the line-of-sight integration point of view, no major changes were
required as the first release  was already designed to deal with non-spherical halo integrations
(in the function {\tt los\_integral()}), although this capability was
not used at the time. 
Triaxiality only requires a few new functions: 
\begin{itemize}
\item {\tt get\_riso\_triaxial()} to transforms $(l,\alpha,\beta)$
integration position into $(x,y,z)$ DM halo coordinates---and if required $(X,Y,Z)$ Euler rotated target
coordinates---from which $R_{\rm iso}$ given in Eq.~(\ref{eq:Riso}) is
calculated; 
\item {\tt mass\_triaxialhalo()} must be used
instead of {\tt mass\_singlehalo()} to obtain the mass of a triaxial
halo. 
\end{itemize}
These changes are transparent
to the user, whose only concern must be to provide the seven triaxiality
initialisation parameters.

\begin{table*}
\caption{\clumpy{} parameters for the user-defined input
  parameter file ({\tt clumpy\_params.txt)}. For the sake
  of completeness, all parameters are reproduced sorted by block
  (cosmology, dark matter, particle physics, etc.), with new parameters
  highlighted in grey, and deprecated ones in strikethrough.}
\vspace{-0.5cm}
\begin{center}
{\small
\begin{tabular}{ll}  \hline\hline
Name & Definition\\ \hline
\multicolumn{2}{l}{{\bf Cosmological parameters} (updated from Planck results)}\\
{\tt gCOSMO\_HUBBLE} & Hubble expansion rate $h=H_0/(100\;{\rm km\;s}^{-1}\;{\rm Mpc}^{-1})$ [-]\\
{\tt gCOSMO\_RHO0\_C} & Critical density of the universe [$M_\odot$~kpc$^{-3}$]\\
{\tt gCOSMO\_OMEGA0\_M} & Present-day pressure-less matter density\\
{\tt gCOSMO\_OMEGA0\_LAMBDA }& Present-day dark energy density\\
&\\[-0.3cm]
\multicolumn{2}{l}{\bf Dark matter parameters}\\\rowcolor{lightgray}
{\tt gDM\_FLAG\_CVIR\_DIST}        & Distribution around $\bar{c}(M)$ from which concentrations are drawn: {\tt\{kLOGNORM, kDIRAC\}}\\\rowcolor{lightgray}
{\tt gDM\_LOGCVIR\_STDDEV}         & Width of log-normal $c(M)$ distribution (if {\tt gDM\_FLAG\_CVIR\_DIST=kLOGNORM})\\\rowcolor{lightgray}
{\tt gDM\_SUBS\_NUMBEROFLEVELS}                & Number of levels for subhaloes\\
{\tt gDM\_MMIN\_SUBS}              & Minimal mass of DM haloes [$M_\odot$] \\
{\tt gDM\_MMAXFRAC\_SUBS}          & Defines the maximal mass of clump in host halo: $M_{\rm max} =~${\tt gDM\_MMAXFRAC\_SUBS}\;$\times\;M_{\rm host}$  \\
{\tt gDM\_RHOSAT}                  & Saturation density for DM [$M_\odot$~kpc$^{-3}$]\\
&\\[-0.3cm]
\multicolumn{2}{l}{{\bf Generic (sub-)halo structural parameters} ({\tt TYPE = DSPH, GALAXY} or {\tt CLUSTER})}\\
{\tt gTYPE\_CLUMPS\_\{FLAG\_PROFILE, \dots\}}     & Description of subhaloes for host {\tt TYPE}: $c(M)$, inner profile, shape parameters \\
{\tt gTYPE\_DPDM\_SLOPE}                          &  Slope of the clump mass function \\
{\tt gTYPE\_DPDV\_\{FLAG\_PROFILE, RSCALE, \dots\}}\!\!\!\! & Spatial distribution of substructures in object {\tt TYPE} \\
{\tt gTYPE\_SUBS\_MASSFRACTION}                   &  Mass fraction of the host halo in clumps \\
&\\[-0.3cm]
\multicolumn{2}{l}{{\bf Milky-Way DM (sub-)halo structural parameters}}\\
{\tt gGAL\_CLUMPS\_\{FLAG\_PROFILE, \dots\}}       & Description of Milky-way DM subhaloes \\
{\tt gGAL\_DPDM\_SLOPE}                            & Slope of clump mass function \\
{\tt gGAL\_DPDV\_\{FLAG\_PROFILE, RSCALE, \dots\}} & Spatial distribution of substructures in object {\tt TYPE} \\
{\tt gGAL\_SUBS\_\{M1, M2, N\_INM1M2\}}            & Number of Milky-Way subhaloes in $[M_1,M_2]$ \\
{\tt gGAL\_\{RHOSOL, RSOL, RVIR\}}                 & Local DM density [GeV~cm$^{-3}$], distance GC--Sun [kpc], virial radius [kpc]\\
{\tt gGAL\_TOT\_\{FLAG\_PROFILE, RSCALE, \dots\}}  & Description of the total DM profile \\\rowcolor{lightgray}
{\tt gGAL\_TRIAXIAL\_AXES[0-3]}                    & Dimensionless
major ($a$), intermediate ($b$), and minor ($c$) axes (see Eq.~(\ref{eq:Riso}))\\\rowcolor{lightgray}
{\tt gGAL\_TRIAXIAL\_ROTANGLES[0-3]}               & Euler rotation angles for triaxial Milky-Way halo [deg] \\\rowcolor{lightgray}
{\tt gGAL\_TRIAXIAL\_IS}                           & Switch-on or off
triaxiality calculation (i.e., use or not the 2 parameters above)\\
&\\[-0.3cm]
\multicolumn{2}{l}{{\bf Particle physics ingredients} (for $\gamma$-ray and $\nu$ flux calculation)}\\\rowcolor{lightgray}
{\tt  gPP\_BR[gN\_PP\_BR]}                     & List of comma-separated values of branching ratios for the 28 channels \\\rowcolor{lightgray}
{\tt  gPP\_DM\_ANNIHIL\_DELTA}                 & For annihilating DM, factor 2 in calculation if Majorana, 4 if Dirac\\\rowcolor{lightgray}
{\tt  gPP\_DM\_ANNIHIL\_SIGMAV\_CM3PERS}       & For annihilating DM, velocity averaged cross-section $\langle\sigma v\rangle_0$ [cm$^3$~s$^{-1}$]\\\rowcolor{lightgray}
{\tt  gPP\_DM\_DECAY\_LIFETIME\_S}             & For decaying DM, lifetime $\tau_{\rm DM}$ of DM candidate [s]\\\rowcolor{lightgray}
{\tt  gPP\_DM\_IS\_ANNIHIL\_OR\_DECAY}         & Switch for annihilating or decaying DM ({\em replace deprecated} \sout{\tt gSIMU\_IS\_ANNIHIL\_OR\_DECAY})\\\rowcolor{lightgray}
{\tt  gPP\_DM\_MASS\_GEV}                      & Mass $m_{\rm DM}$ of the DM candidate [GeV] \\\rowcolor{lightgray}
{\tt  gPP\_FLAG\_SPECTRUMMODEL}                & Model to calculate final state ({\em replace deprecated} {\sout{\tt gDM\_GAMMARAY\_FLAG\_SPECTRUM}})\\\rowcolor{lightgray}
{\tt  gPP\_NUMIXING\_THETA\{12, 13, 23\}\_DEG} & Neutrino mixing angles [deg]\\
&\\[-0.3cm]
\multicolumn{2}{l}{{\bf Simulation parameters/outputs} (for a given \clumpy{} run)}\\
{\tt gLIST\_HALOES}                            & DM haloes considered in $J$-factor calculations [default={\tt data/list\_generic.txt}]\\\rowcolor{lightgray}
{\tt gLIST\_HALOES\_JEANS}                     & Objects considered in Jeans's analysis [default={\tt data/list\_generic\_jeans.txt}]\\
{\tt gSIMU\_ALPHAINT\_DEG}                     & Integration angle $\alpha_{\rm int}$ [deg] (if {\tt gSIMU\_HEALPIX\_NSIDE} not -1, use HEALPix resolution)\\
{\tt gSIMU\_EPS}                               & Precision used for any operation requiring one (numerical integration, \dots)\\
{\tt gSIMU\_SEED}                              & Seed of random number generator to draw clumps (if 0, from computer clock)\\\rowcolor{lightgray}
{\tt gSIMU\_FLAG\_NUFLAVOUR}                   & Choice of neutrino flavour ({\tt kNUE, kNUMU, kNUTAU})\\\rowcolor{lightgray}
{\tt gSIMU\_FLUX\_AT\_E\_GEV}                  & Energy (GeV) at which to calculate fluxes  \\\rowcolor{lightgray}
{\tt gSIMU\_FLUX\_E\_MIN}                      & Lower energy bound (GeV) for the integrated flux calculation\\\rowcolor{lightgray}
{\tt gSIMU\_FLUX\_E\_MAX}                      & Upper energy bound (GeV) for the integrated flux calculation\\\rowcolor{lightgray}
{\tt gSIMU\_GAUSSBEAM\_GAMMA\_FWHM\_DEG}       & Gaussian beam [deg] for $\gamma$-ray detector for skymaps smoothing (no smoothing if set to -1)\\\rowcolor{lightgray}
{\tt gSIMU\_GAUSSBEAM\_NEUTRINO\_FWHM\_DEG}    & Gaussian beam [deg] for $\nu$ detector for skymaps smoothing (no smoothing if set to -1)\\\rowcolor{lightgray}
{\tt gSIMU\_HEALPIX\_NSIDE}                    & $N_{\rm side}$ of
HEALPix maps (if -1, set to be as close as possible to $\alpha_{\rm int}$) \\\rowcolor{lightgray}
{\tt gSIMU\_HEALPIX\_RING\_WEIGHTS\_DIR}       & Ring weights directory for improved quadrature (optional)\\\rowcolor{lightgray}
{\tt gSIMU\_IS\_ASTRO\_OR\_PP\_UNITS}          & Outputs (plots and files) in astro ($M_\odot$ and kpc) or particle physics (GeV and cm) units. \\\rowcolor{lightgray}
{\tt gSIMU\_IS\_WRITE\_FLUXMAPS}               & For 2D skymaps, whether to save or not $\gamma$-ray and $\nu$ fluxes (the $J$ factor is always saved) \\\rowcolor{lightgray}
{\tt gSIMU\_IS\_WRITE\_FLUXMAPS\_INTEG\_OR\_DIFF}\!\!\!\!& If {\tt gSIMU\_IS\_WRITE\_FLUXMAPS} is true, whether to save integrated or differential fluxes\\\rowcolor{lightgray}
{\tt gSIMU\_IS\_WRITE\_GALPOWERSPECTRUM}       & Whether to calculate (and save) or not the DM power-spectrum for the Milky-Way \\\rowcolor{lightgray}
{\tt gSIMU\_IS\_WRITE\_ROOTFILES}              & Whether to save or not {\tt .root} files even if option -p is used (not enabled for skymaps and 'stat') \\\rowcolor{lightgray}
{\tt gSIMU\_OUTPUT\_DIR}                       & Output directory to select other than local run (directory is {\tt output/} if set to -1) \\
%
%&\\[-0.25cm]
%
\hline
\end{tabular}
}
\label{tab:param}
\end{center}
\vspace{-0.5cm}
\end{table*}

%%%%%%%%%%%%%%%%%%%%%%%%%%%%%%%%%%%%%%%%%%%%%%%%%%%%%%%%%%%%%%%%%%%%%%%%%%%%%%
%%%%%%%%%%%%%%%%%%%%%%%%%%%%%%%%%%%%%%%%%%%%%%%%%%%%%%%%%%%%%%%%%%%%%%%%%%%%%%
\section{New outputs (content and format) and pixelisation \label{sec:outputs}}

\subsection{$\gamma$-ray and $\nu$ spectra: {\tt spectra.h} \label{sec:gammaray_nu}}
We add the gamma and neutrino spectra from dark matter annihilation
and decay. We use the tabulated spectra of recent PYTHIA simulations \cite{2011JCAP...03..051C,2014JCAP...03..053B}
including or not EW corrections \cite{2011JCAP...03..019C}, in which a Higgs mass of 125 GeV is assumed.

\subsubsection{Implementation of PPPC4DMID in \clumpy{}}
The values are calculated by a 2D linear interpolation on $\log(E)$ and
$m_{\rm DM}$ from tabulated spectra \cite{2011JCAP...03..019C}.

\begin{itemize}
  \item Branching ratios: {\tt gPP\_BRANCHINGRATIO\_LIST} (see table~\ref{tab:param})
  is a list of comma-separated values for the branching ratios\footnote{For a minimum of safety, $\sum_{i=1}^{28}BR_{i}\leq1$
%$\sum_{i=1}^{28}BR_{i}\leq1 -(BR_{W_L^+W_L^-}+BR_{W_T^+W_T^-}+BR_{Z_LZ_L}+BR_{Z_TZ_T}$
 is checked and indicated to the user but $\sum_{i=1}^{28}BR_{i}>1$
 can be allowed due to the possible redundancy between channels with polarisations or chiralities.} of the 28 primary channels
$e_{L}^+e_L^-$, $e_R^+e_R^-$, $e^+e^-$, $\mu_L^+\mu_L^-$, $\mu_R^+\mu_R^-$, $\mu^+\mu^-$,
$\tau_L^+\tau_L^-$, $\tau_R^+\tau_R^-$, $\tau^+\tau^-$,
$q\bar{q}$, $c\bar{c}$, $b\bar{b}$, $t\bar{t}$, $W_L^+W_L^-$, $W_T^+W_T^-$, $W^+W^-$, $Z_LZ_L$, $Z_TZ_T$, $ZZ$,
$gg$, $\gamma\gamma$, $hh$, $\nu_e\nu_e$, $\nu_{\mu}\nu_{\mu}$, $\nu_{\tau}\nu_{\tau}$, $VV
\rightarrow 4e$, $VV \rightarrow 4\mu$, $VV \rightarrow 4\tau$.

  \item Final state: {\tt gENUM\_FINALSTATE} (see table~\ref{tab:enum}) must be chosen
  among {\tt kGAMMA} ($\gamma$-rays) and {\tt kNEUTRINO} ($\nu$). The flavour of the latter is chosen
  (see table~\ref{tab:param}) among {\tt gSIMU\_FLAG\_NUFLAVOUR = kNUE, kNUMU, kNUTAU}.
  
  \item Decay or annihilation: it is set by the boolean {\tt gPP\_DM\_IS\_ANNIHIL\_OR\_DECAY}. For a decaying DM particle, we use the same spectra $dN^f_{\gamma,\nu}/dE$ as for the case of annihilation, but assuming $dN^{f, {\rm decay}}_{\gamma,\nu}/dE(m_{\rm DM}) = dN^{f,{\rm ann}}_{\gamma,\nu}/dE(m_{\rm DM} / 2)$.
\end{itemize}

\subsubsection{Neutrino mixing}
The simulations from \cite{2011JCAP...03..051C} provide neutrino production
spectra. For distant astrophysical sources, the journey in
vacuum and transition between the different flavour states can be
described by average oscillations \cite{1987RvMP...59..671B}:
\beq
P(\nu_l\rightarrow\nu_{l'})=P(\bar{\nu_l}\rightarrow\bar{\nu}_{l'})=\sum_{k=1}^3
|U_{l'k}|^2 |U_{lk}|^2
\label{eq:nuoscillations}
\eeq
where $U$ is the neutrino mixing matrix and $k=1,2,3$ for the 3 mass eigenstates.
The oscillation matrix is filled with {\tt nu\_oscillationmatrix()} from
the mixing angles $\{\theta_{12},\,\theta_{23},\,\theta_{13}\}$
({\tt gPP\_NUMIXING\_THETA\{12,13,32\}\_DEG} (see table~\ref{tab:param}),
whose default values are taken from \cite{2014PhRvD..90i3006F},
i.e. $\{34^\circ,\, 49^\circ,\,9^\circ\}$. 
The $CP$ phase is taken to be zero (see \cite{2014PhRvD..90i3006F} for a recent discussion).
Oscillation effects are applied to the $\nu_e,\nu_{\mu}$ and $\nu_{\tau}$
spectra. The code gives the resulting $\nu_{e}$, $\nu_{\mu}$ or
$\nu_{\tau}$ fluxes\footnote{Relevant quantities for DM detection with neutrino telescopes
are $\nu_{\mu}+\bar{\nu}_{\mu}= 2\times \nu_{\mu}$. This factor 2 for anti-neutrino
contribution is not accounted for in \clumpy{} results.}.

\subsubsection{Particle physics term and flux in \clumpy{}}
The above final states, the particle physics term Eq.~(\ref{eq:term-pp}),
or the $\gamma$-ray (or neutrino) flux (Eq.~\ref{eq:flux-general}) for a
given astrophysical factor can be displayed with {\tt ./bin/clumpy -z}.
The corresponding functions in \clumpy{} are {\tt dNdE()},
{\tt dPPdE()}, {\tt dPPdE\_integrated()}, and {\tt flux()}.
Fluxes (and integrated fluxes) are also displayed whenever the astrophysical
factor for the Galaxy ({\tt ./bin/clumpy -g}) or for a DM halo ({\tt ./bin/clumpy -h})
is calculated.

\subsection{Map handling and outputs: {\tt healpix\_fits.h}}
This new release provides additional tools in the context of 2D maps,
as well as more advanced output options in addition to the {\tt ASCII} files outputs and plots which were available with the first version.
\subsubsection{\healpix{} map tools}
\clumpy{} is now interfaced with the 
\healpix{} ({\sc H}ierarchical {\sc E}qual
{\sc A}rea iso{\sc L}atitude {\sc Pix}elation) package, which provides a large set of routines for efficient manipulation and
analysis of numerical data on the discretized sphere \citep{Gorski2005}. Within the 
\healpix{} framework, \clumpy{} now robustly handles large field of views (FOV)
at arbitrary positions of the sky, and the fast calculation of full skymaps.

Various shapes for the part-sky grid 
are now available (i.e., circular, great-circle or iso-latitude/longitude rectangular shapes of the FOV).
Reversely, for full-sky calculations, simple masking shapes can be applied to reduce
computation time, memory usage and output file size of the simulation. Examples
for available FOV choices and masks are given in the \doxygen{} documentation.

Using the \healpix{} library routines, \clumpy{} now
provides the options of smoothing the output $J$-factor skymaps
with a Gaussian beam and calculating the angular power spectrum
(APS) of the maps. Up to two smoothing kernels may be specified (e.g., one for a $\gamma$-ray
instrument and one of a neutrino observatory), both being applied on
the same output map originally computed by \clumpy{}\footnote{In
order to avoid smoothing the edges of part-sky or masked maps, \clumpy{}
appropriately extends the original grid when smoothing is
  requested by the user.}.
The APS calculation\footnote{Note that power spectra for part-sky and masked
maps are affected by power suppression and spectral leakage effects according
to the shape of the windowing function of the FOV. Further accounting and
correction for these effects must be performed outside of \clumpy{}.} is limited to the Galactic mode 
({\tt ./bin/clumpy -g}) and is performed independently for the
total $J$-factor  profile of the halo $J_{\rm tot}$, the smooth halo contribution
$J_{\rm sm}$, the substructure component $J_{\rm subs}$ (average $\langle J_{\rm subs}\rangle$ and drawn haloes $J_{\rm drawn}$ contributions) and the cross-product $J_{\rm cross-prod}$ (annihilations only)\footnote{Note that the power spectrum $C_\ell$ of the
  $J$-factor components is an additive quantity, $C_\ell^{\rm tot}=
  C_\ell^{\rm sm} + C_\ell^{\rm subs} + \ldots$\;\;.}.

\subsubsection{Output format}
The output files of the 2D skymaps and APS are now written in the \fits{} (Flexible
Image Transport System) file format \citep{Pence2010}. This format allows to
clearly and efficiently store several maps as binary
tables in one single file, and to append extensive meta information in
the \fits{} headers of the output file. Both full-sky and part-sky \healpix{} maps can be
handled in the \fits{} format. \clumpy{} \fits{} files are composed of
several extensions ($J$-factor skymaps, smoothed $J$-factor skymaps, fluxes), each containing different contributions ($J_{\rm sm}$, $J_{\rm subs}$,
$J_{\rm cross-prod}$,\ldots) in separate columns of the extension. Each
extension possesses its own \fits{} header, containing the relevant input
parameters of the performed simulation. Writing \fits{} file in \clumpy{} is done by using the \cfitsio{}
library either
directly, or indirectly via existing \healpix{} routines.

Many programs are available for reading, displaying and
manipulating \healpix{} maps in the \fits{} format\footnote{\url{http://fits.gsfc.nasa.gov/fits_viewer.html}}
(e.g. fv\footnote{\url{http://heasarc.gsfc.nasa.gov/ftools/fv/}}, DS9\footnote{\url{http://ds9.si.edu}},
Aladin\footnote{\url{http://aladin.u-strasbg.fr}}, Healpy from the \healpix{} package, \dots). 
To facilitate the display and post-processing of 2D-simulations, the {\tt
  ./bin/clumpy -o} mode has been implemented to convert and extract the output maps
in different file formats (more details are given in the {\tt README} \clumpy{} file).
In particular, {\tt ASCII} files similar to those
of the first \clumpy{} release can be
written. For 2D \rootcern{} plots and output, the data is transformed from the \healpix{} pixelisation scheme onto a rectangular grid in 
longitudinal and latitude coordinates. This tangential plane approximation is only appropriate for small FOV.

%%%%%%%%%%%%%%%%%%%%%%%%%%%%%%%%%%%%%%%%%%%%%%%%%%%%%%%%%%%%%%%%%%%%%%%%%%%%%%
%%%%%%%%%%%%%%%%%%%%%%%%%%%%%%%%%%%%%%%%%%%%%%%%%%%%%%%%%%%%%%%%%%%%%%%%%%%%%%
\section{New science: Jeans analysis\label{sec:jeans}}

The original goal of the \clumpy{} code was to provide a robust and versatile
tool to compute the astrophysical factors for DM annihilation and
decay signals. The updates and new features introduced to improve \clumpy{} in that
respect were presented in \S\ref{sec:updatesJ}. Here, we move to
an entirely new aspect of this release, namely the
possibility to run a Jeans analysis coupled to an MCMC engine, in order to
derive data-driven DM density profiles from a set of stellar/galactic kinematic
data. This method is independent from the $J$-factor calculation itself. The Jeans module presented here has already been used in \cite{2015MNRAS.446.3002B},
\cite{2015MNRAS.453..849B} and \cite{2015ApJ...808L..36B}, in order to provide robust $J$-factors (and
associated error bars) for both the `classical' and `ultrafaint' dwarf spheroidal galaxies of the
Milky Way, which are among the most promising targets for DM indirect detection. The recently discovered
dSph galaxy candidates \cite{2015ApJ...805..130K,2015ApJ...807...50B} are typical objects on which
to apply the Jeans analysis once kinematics data become available, in order to obtain their $J$-factor.

          %%%%%%%%%%%%%%%%%%%%%%%%%%%%%%%%%%
\subsection{Jeans equations}
\label{subsec:jeans_theo}
The Jeans equation is obtained after integrating the
collisionless Boltzmann equation in spherical symmetry, assuming steady-state and negligible rotational support \citep{2008gady.book.....B}.
 It relates the dynamics of a collisionless tracer population
(e.g. stars in a dwarf spheroidal galaxy or galaxies in a galaxy cluster) to the underlying gravitational
potential. Namely, the Jeans equation reads
\beq
\frac{1}{\nu}\frac{d}{dr}\left(\nu\bar{v_r^2}\right)+2\frac{\beta_{\rm
    ani}(r)\bar{v_r^2}}{r} = -\frac{GM(r)}{r^2}\;,
\eeq 
with 
\beq
M(r)=4\pi\int_{0}^{r} \rho_{\rm tot}(s) s^2 ds\;,
\eeq
and where:
\begin{itemize}
\item the definition of the enclosed mass $M(r)$ assumes that the tracer
  population (i.e. the stars) has a negligible mass compared to the
  underlying DM halo. The density profile of the latter is $\rho_{\rm
    tot}(r)$;
\item $\nu(r)$ is the 3D number density (or light profile) of the tracer population. Its 2D projection is called surface brightness (see below);
\item $\bar{v_r^2}$ is the radial velocity dispersion of the tracers;
\item $\beta_{\rm ani}(r) \equiv 1 - \bar{v_\theta^2}/\bar{v_r^2}$ is
  the velocity anisotropy of the tracers, and depends on the ratio of the tangential to
  the radial velocity dispersions.
\end{itemize}
The formal solution to the 3D Jeans equation is
\beq
\nu(r)\bar{v_r^2}(r)=\frac{1}{f(r)}\times \int_r^{\infty}
f(s)\nu(s)\frac{GM(s)}{s^2} ds\;, 
\eeq
with
\beq 
f(r) = f_{r_1} \exp \left[\int_{r_1}^{r}\frac{2}{t}\beta_{\rm ani}(t)dt \right].
\label{f}
\eeq
In practice, observations provide only the 2D projections on the sky of the tracer velocity dispersion and number density. For a given projected radius $R$, the projected 2D solution of the
Jeans equation is
\beq
I(R)\sigma_p^2(R) = 2\int_0^\infty \left(1-\beta_{\rm ani}(r)\frac{R^2}{r^2}\right)\nu(r)\bar{v_r^2}(r)dr\;,
\label{eq:jeans_proj}
\eeq
with $I(R)$ and $\sigma_p(R)$ the surface brightness and
projected velocity dispersion, respectively.
A review on recent developments and applications of the Jeans analysis
can be found, e.g., in  \cite{2014RvMP...86...47C}.

          %%%%%%%%%%%%%%%%%%%%%%%%%%%%%%%%%%
\subsection{Implementation in the new library {\tt jeans\_analysis.h}}
\label{sec:jeans_great}

Computing Eq.~(\ref{eq:jeans_proj}) requires three levels of
imbricated integrations which can be time consuming. For some specific
parametrisations of $\beta_{\rm ani}(r)$, analytical shortcuts may be
found to reduce the problem to a single integration. These shortcuts,
based on the kernel implementation of Eq.~(\ref{eq:jeans_proj}) of
\cite{2005MNRAS.363..705M}, are implemented in \clumpy{}, along with the full numerical
integration whenever necessary. This approach is fully described in
the documentation attached to the {\tt jeans\_analysis.h} library 
and is not detailed again in this paper. 

The functions at the core of the library are related to the description of the ingredients of the Jeans analysis:
\begin{itemize}
\item {\tt beta\_anisotropy(gENUM\_ANISOTROPYPROFILE,\dots)}: returns the velocity anisotropy profile
  selected by the user in the file containing the halo structural parameter
  description. The most generic form for the anisotropy is the Baes \& Van Hese
  parametrisation~\cite{2007A&A...471..419B},
\beq
\beta_{\rm ani}(r)=\frac{\beta_0+\beta_\infty (r/r_a)^\eta}{1+(r/r_a)^\eta}\;,
\eeq 
with four free parameters ($\beta_0$, $\beta_\infty$, $r_a$ and
$\eta$). The user may also choose the constant parametrisation or the Osipkov-Meritt
parametrisation~\cite{1979PAZh....5...77O,1985AJ.....90.1027M}, which both are special limiting cases of the Baes \& Van
Hese profile (see Table~\ref{tab:enum}).
\item {\tt light\_profile(gENUM\_LIGHTPROFILE,\dots)}: returns the 2D or 3D light
  profile (i.e. $I(R)$ or $\nu(r)$) selected by the user in the file containing
  the light profile structural parameters. Four 2D profiles and two 3D profiles
  are available (see Table~\ref{tab:enum}). 
  Depending whether the user selects a 2D or a 3D light profile, the corresponding
  (de-)projections are performed by \clumpy{} in order to solve the Jeans equation.
\item {\tt jeans\_*}: a series of functions which provide the necessary
  steps to effectively solve the Jeans equation, through the various
  integrations detailed in \S\ref{subsec:jeans_theo}.
\end{itemize}

          %%%%%%%%%%%%%%%%%%%%%%%%%%%%%%%%%%
\subsection{Jeans analysis within \clumpy{}: inputs and outputs}
The standard procedure of a Jeans analysis is the following:
\begin{enumerate}
\item fit the surface brightness profile $I(R)$ with a given parametric function\footnote{While \clumpy{} does not perform the actual fit to the surface brightness profile, several parametrisations of $I(R)$ are implemented in the code as to perform the Jeans analysis (see {\tt gENUM\_LIGHTPROFILE} enumerator in Table~\ref{tab:enum}).}; 
\item choose a parametrisation for the DM density profile $\rho_{\rm tot}(r)$ and the tracer anisotropy $\beta_{\rm ani}(r)$;
\item compute the projected velocity dispersion $\sigma_p$ from $I(R)$, $\rho_{\rm tot}(r)$ and $\beta_{\rm ani}(r)$ (Eq. \ref{eq:jeans_proj}); 
\item find the DM density and anisotropy parameters that best reproduce the kinematic observables.
\end{enumerate}

\begin{figure}[!t]
\centering
\includegraphics[width=\columnwidth]{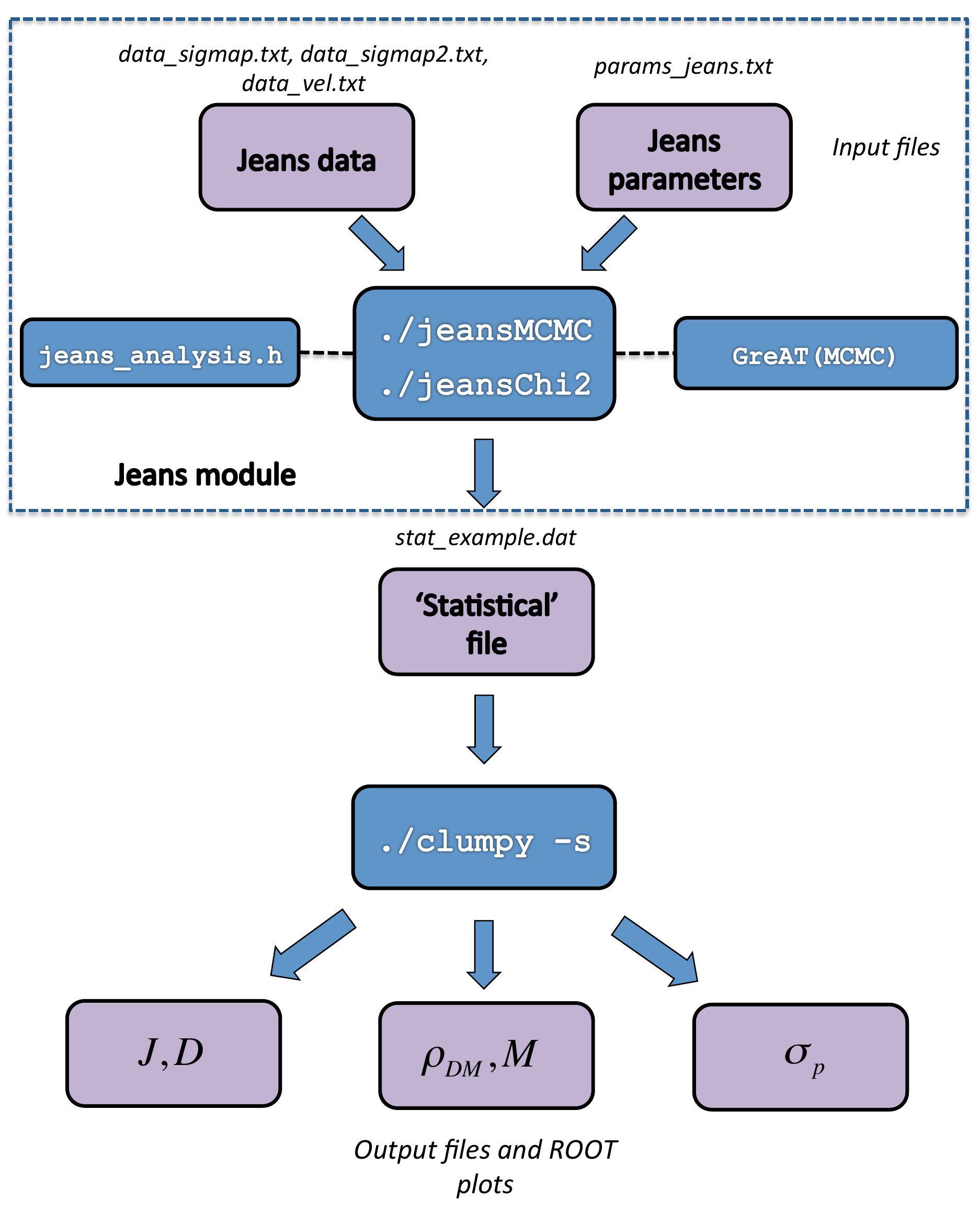}
\caption{Diagram of the Jeans analysis with \clumpy{}. From a kinematic data file and a parameter file describing the free parameters, a statistical Jeans analysis can be run with {\tt ./jeansMCMC} or {\tt ./jeansChi2}. A `statistical' output file is created, from which estimators of different quantities of interest (i.e., $J$-factors) can be obtained with {\tt ./clumpy -s}.}
\label{fig:jeans_analysis}
\end{figure}

The DM density profile can then be used to compute the $J$-factor, or any other DM-related quantity. Figure \ref{fig:jeans_analysis} summarises the steps of the Jeans analysis, and the related executables. We describe below the different ingredients of the analysis.

\subsubsection{Input files: kinematic data and free parameters}

\paragraph{Kinematic data} The kinematic data of the tracer population
(i.e., stars in a dwarf spheroidal galaxy) are usually in the form of
velocity dispersion profiles $\sigma_{\rm obs}$ (resp. squared
velocity dispersion, $\sigma_{\rm obs}^{2}$), or line-of-sight
velocities $v_i$. These data must be filled in specific data
files. Three examples are given in the {\tt data/} directory, for the
three different types of kinematic data: {\tt data\_sigmap.txt}, {\tt
  data\_sigmap2.txt}, and {\tt data\_vel.txt}. A keyword is associated
to each data type and written in the header of the file: {\tt Sigmap}, {\tt Sigmap2}, and {\tt Vel}. Note that surface brightness profiles can also be read by \clumpy{} (see the example file {\tt data\_light.txt}).

\paragraph{Free parameters} 
The surface brightness, velocity anisotropy and DM density profiles used in the Jeans
analysis are set in the parameter file {\tt params\_jeans.txt}, located in the {\tt
/data} directory. Up to nine parameters can be let free in the
analysis to describe velocity anisotropy and DM density profiles. Their ranges are set
in {\tt params\_jeans.txt}. The parameters describing the surface brightness profile,
as well as some characteristics of the object (e.g. name, distance, ...), must also be
filled in this parameter file.

\subsubsection{Likelihood functions}
In {\tt jeans\_analysis.h}, we provide two likelihood functions {\tt log\_likelihood\_jeans\_\{binned,unbinned\}()} for the fit of the velocity dispersion $\sigma_p$ to the kinematic observables. If the kinematic data are binned velocity dispersion profiles, the likelihood function is\footnote{If the kinematic data are binned \textit{squared} velocity dispersion profiles, then $\sigma$ is replaced by $\sigma^2$ in Eq. (\ref{eq:likelihood_binned})}:
\begin{equation}
  \mathcal{L}^{\rm bin}= \prod_{i=1}^{N_{\rm bins}}
  \frac{(2\pi)^{-1/2}}{\Delta{\sigma_i}(R_i)} \exp\biggl
         [-\frac{1}{2}\biggl (\frac{\sigma_{\rm obs}(R_{i})
             -\sigma_\text{p}(R_i)}{\Delta {\sigma}_i(R_i)}\biggr
           )^{2}\biggr ],
  \label{eq:likelihood_binned}
\end{equation}
with $\Delta \sigma_i(R_i)$  the error on the velocity dispersion at the radius $R_i$. If the data are in the form of line-of-sight individual velocities $v_i$, then the likelihood function used is \citep{2008ApJ...678..614S}:
\begin{equation}
  \mathcal{L}^{\rm unbin}=\!\prod_{i=1}^{N_{\rm tracers}}\!
  \left(\frac{(2\pi)^{-1/2}}{\sqrt{\sigma_\text{p}^2(R_i)+\Delta_{v_{
          i}}^{2}}}\! \exp\biggl [\!-\frac{1}{2}\biggl (\frac{(v_{\rm
        i}
      -\bar{v})^{2}}{\sigma_\text{p}^2(R_i)+\Delta_{v_{i}}^{2}}\biggr
    )\biggr ]\right)^{P_i},
  \label{eq:likelihood_unbinned}
\end{equation}
with $P_i$ the membership probability of the $i$-th tracer\footnote{$P_i$ estimates the probability that the $i$-th tracer belongs to the object. This quantity is often available for stars in dwarf spheroidal galaxies.}. The underlying assumption is that the line-of-sight velocity distribution is Gaussian, centred on the mean velocity of the tracers $\bar{v}$, with a dispersion of velocities at radius $R_i$ of the $i$-th tracer coming from both the intrinsic dispersion $\sigma_\text{p}(R_i)$ from
Eq.~(\ref{eq:jeans_proj}) and the velocity measurement uncertainty $\Delta_{v_{i}}$.

\subsubsection{MCMC and bootstrap analyses}
With \clumpy{}, the Jeans analysis can be done in two ways. The first
uses the MCMC toolkit \great{}, and the second a simple $\chi^2$
minimisation with a bootstrap analysis.
\paragraph{MCMC analysis} 
The MCMC technique, based on Bayesian parameter inference, allows for an efficient sampling of the posterior probability density function (PDF) of a vector of parameters with Markov chains. Credibility intervals (CIs) can then be reconstructed from the PDFs, for any quantity of interest. \clumpy{} is interfaced with the Grenoble Analysis Toolkit (\great{}) \citep{Putze:2014aba}, a public {\tt C++} statistical framework which handles MCMC analysis with the Metropolis-Hastings algorithm \citep{hastings70}. To use \great{} with \clumpy{}, the user has to create the environment variable `\great{}', pointing to the installation folder of {\tt GreAT}. The {\tt jeansMCMC} executable provides an example of MCMC analysis, using the input files and likelihood functions described in the previous sections.
\paragraph{Bootstrap analysis}
The Jeans analysis can also be run with a simple $\chi^2$ minimisation, using the {\tt Minimizer} class of the \rootcern{} libraries. The $\chi^2$ function is
\begin{equation}
\chi^2 = -2 \log(\mathcal{L}),
\end{equation}
with $\mathcal{L}$ being one of the likelihood functions defined in the previous section (Eqs. (\ref{eq:likelihood_binned}) or (\ref{eq:likelihood_unbinned})). The {\tt jeansChi2} executable provides an example of $\chi^2$ analysis.

With the same executable, the user can also run a bootstrap
analysis for an estimation of the statistical error bars
\citep{1982jbor.book.....E}, which proceeds as follows: i) for a
given kinematic data sample of $n$ line-of-sight velocities, $N$
re-samples are generated by drawing with replacement $n$ velocities
among the original sample; ii) for each re-sample, the best-fitting
anisotropy and DM parameters are found using the $\chi^2$
minimisation; iii) the bootstrap estimator of any quantity of interest is the mean over the $N$ configurations, and the dispersion is used as statistical uncertainty.

\subsubsection{`Statistical' output file}
Each Jeans executable creates an output `statistical' file, readable by \clumpy{} with the option {\tt -s}. The latter allows to draw the estimators of several DM-related quantities, such as the astrophysical factors or the density profile, from the results of the statistical analysis (i.e., median and credible intervals from an MCMC analysis, mean and and dispersion for a bootstrap analysis). With respect to the previous \clumpy{} version, new options were added to draw several Jeans-related quantities,
i.e. $\sigma_p(R)$, $I(R)$, etc. 

%%%%%%%%%%%%%%%%%%%%%%%%%%%%%%%%%%%%%%%%%%%%%%%%%%%%%%%%%%%%%%%%%%%%%%%%%%%%%%
\section{Installation and run examples \label{sec:examples}}

In this section, we provide a quick description of \clumpy{}'s architecture and installation, the various options for running the code and, finally, a few run and plot examples obtained with \clumpy{}.
Many more examples and illustrations are provided along with the \doxygen{} documentation.

\subsection{Code installation/compilation/executables\label{sec:code}}

\clumpy{} is written in C/C++ and relies of the \rootcern{},
\cfitsio{}, \healpix{} (C++ and F90), and \gsl{} libraries.
Instructions for installations and links to downloads are given in the {\tt README} file.
The code's structure is standard, with separate directories for the various pieces
of code: declarations are in {\tt include/*.h}, sources in {\tt src/*.cc}, compiled libraries,
objects and executables are respectively in the {\tt lib/}, {\tt obj/}, and {\tt bin/} directories. 
With respect to the first version, two new libraries were created ({\tt jeans\_analysis.h}
and {\tt healpix\_fits.h}), one greatly expanded ({\tt spectra.h}), and two new
executables added for the Jeans analysis ({\tt jeansChi2.cc} and {\tt jeansMCMC.cc}). 

The various executables (and options) available for a Jeans or $J$-factor related
\clumpy{} run should be self-explaining (just type the command lines below).
See the next section for examples.

\subsubsection{Jeans analysis}
As detailed above (\ref{sec:jeans_great}), the Jeans analysis allows to extract the DM density profile of a DM halo from the kinematic data of its tracer population.
   \begin{itemize}
     \item {\tt ./bin/jeansChi2}: Jeans analysis with simple
       $\chi^2$ minimisation and bootstrap analysis.
     \item {\tt ./bin/jeansMCMC}: Jeans/MCMC analysis (only if \great{} is installed).
   \end{itemize}

\subsubsection{$J$-factor (and flux) analysis } 
The list of available options have not changed much, but they include all the
refinements described in this paper (for the calculation, outputs, and displays).
   \begin{itemize}
     \item {\tt ./bin/clumpy -g[option]}: $J$-factor in Galaxy for smooth and clumps (mean or drawn).
     \item {\tt ./bin/clumpy -h[option]}: $J$-factor in any halo for smooth and clumps (mean or drawn).
     \item {\tt ./bin/clumpy -o[option]}: \clumpy{} skymap file manipulation.
     \item {\tt ./bin/clumpy -s[option]}: PDF and confidence levels on $\rho(r)$, $J$, $\sigma_p(R)$...
     \item {\tt ./bin/clumpy -z}: spectra, particle physics term, and flux (annihilation, decay).
     \item {\tt ./bin/clumpy -f}: append extension with flux maps to existing FITS file.
   \end{itemize}

          %%%%%%%%%%%%%%%%%%%%%%%%%%%%%%%%%%
\subsection{Sky patches, smoothing, $\gamma$-rays and $\nu$\label{subsec:{skymap}}}

\begin{figure}[!t]
\centering
\includegraphics[width=\columnwidth, trim=0 0 0 150 , clip=true]{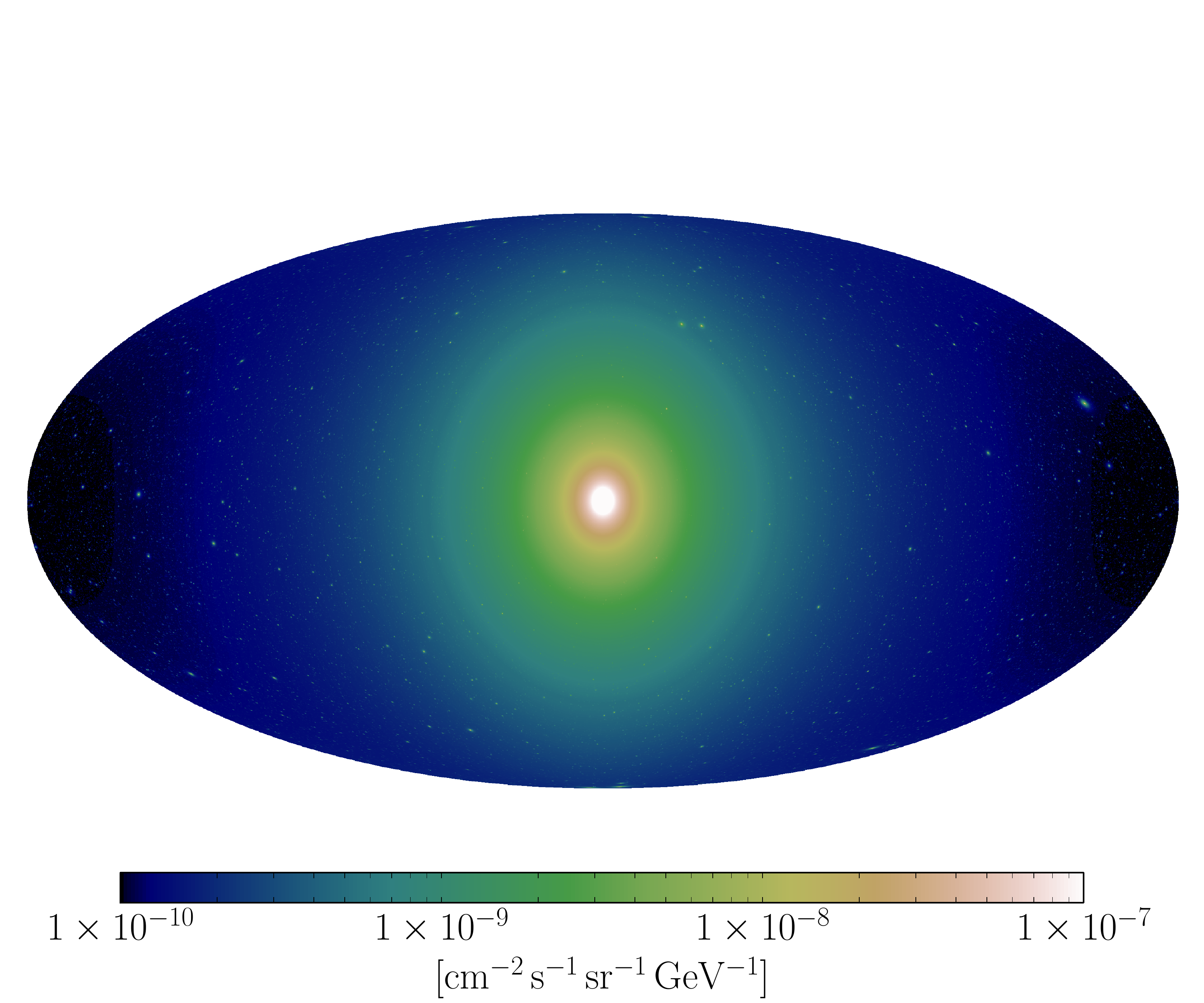}
\includegraphics[width=\columnwidth, trim=0 0 0 100 , clip=true]{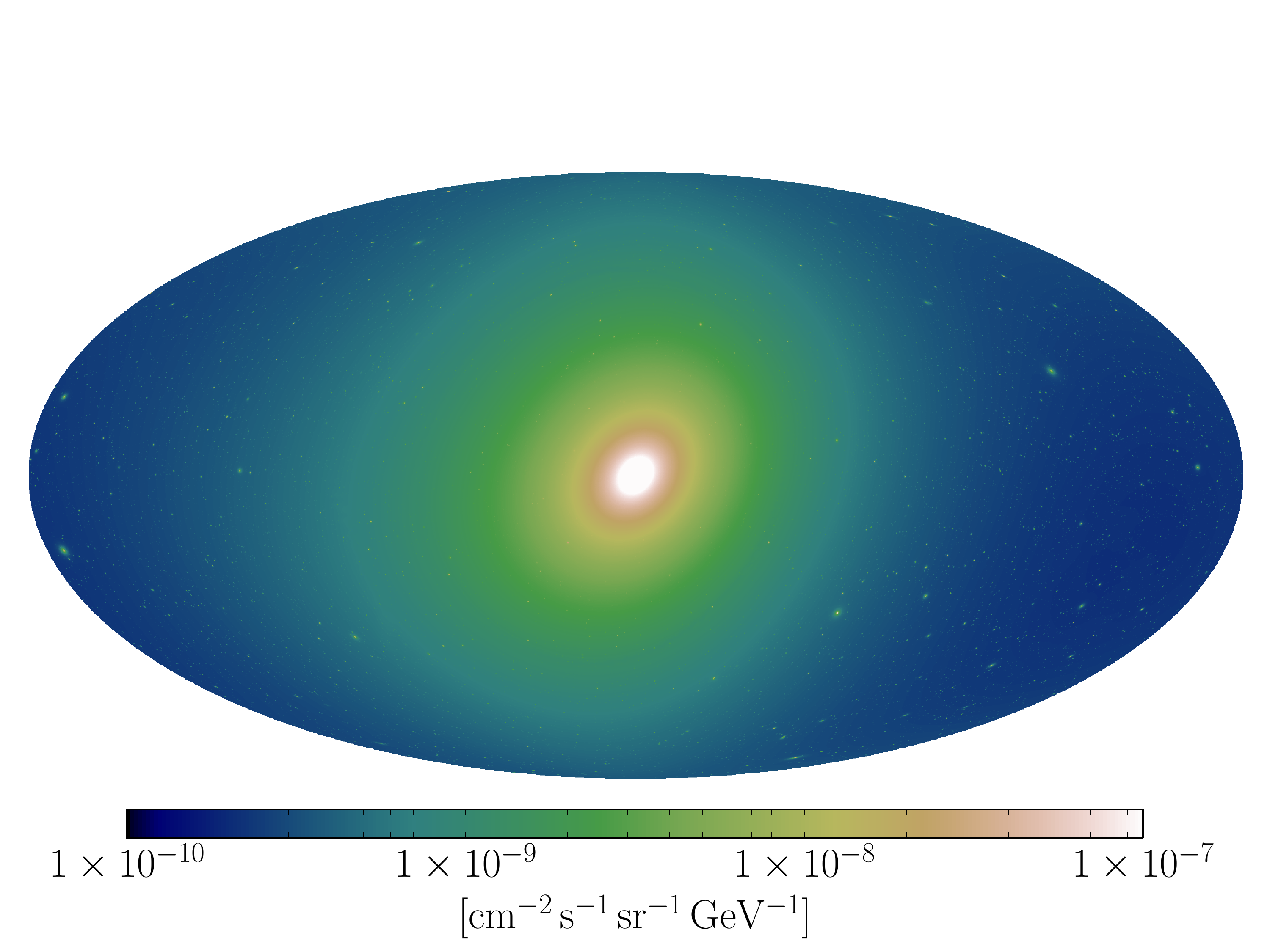}
\caption{{\em Top:} 
Differential intensity skymap of the full galactic halo for {\grs} from annihilation at $4\,\rm{GeV}$ for $m_{\chi}=200\,\rm{GeV}$ and $\chi\chi\rightarrow b\bar{b}$ channel. The skymap is drawn in the {\tt -g7} mode for a numeric resolution of $N_{\rm side} = 2^{9}$ (corresponding to a pixel diameter of $\sim 0.12^{\circ}$) with the parameters for a spherically symmetric halo specified in the text. For a relative error  $RE_{J_{clumps}}\leq 2\%$ (see Eq. 28 in \cite{2012CoPhC.183..656C})  and a numeric precision of $1\%$, a total number of $164,{}186$ substructures is drawn within $7\,h$, consuming $1.0\,GB$ RAM. {\em Bottom:} Galactic halo computed with the same parameters as on the top panel, but with a triaxially distorted shape of the total halo. The semi-axes ratios are, motivated by \cite{Law2009}, set to $b/a = 0.83$, $c/a = 0.67$, and $a = 1.47$. The spheroid is rotated versus the line of sight by the Euler angles $\alpha = 30^{\circ}$, $\beta = 40^{\circ}$ and $\gamma = 20^{\circ}$ (arbitrary values chosen for illustration only). This skymap needs $3\,h$ of computation time ($820\,MB$ RAM) for $36,{}994$ drawn substructures. However, as the spherical symmetry of the halo is now broken, a substantial amount of computation time ($\sim1\,hr$) is required for the computation of the smooth $J-$factor components.}
\hspace{0.08\textwidth}
\label{fig:fullskymaps}
\vspace{-0.5cm}
\end{figure}

\begin{figure}[!t]
\centering
\includegraphics[width=\columnwidth, trim=0 0 0 30 , clip=true]{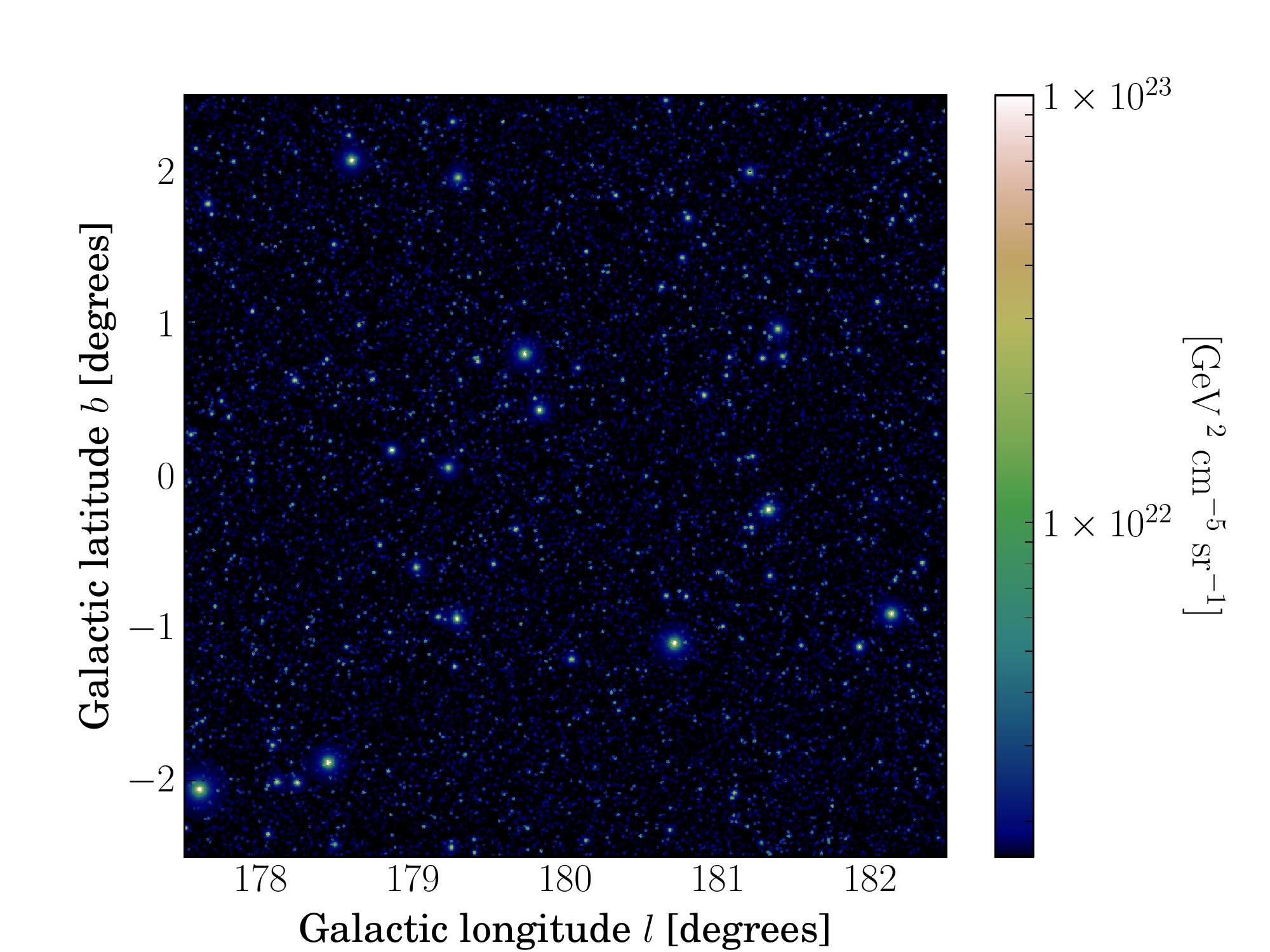}
\includegraphics[width=\columnwidth, trim=0 0 0 30 , clip=true]{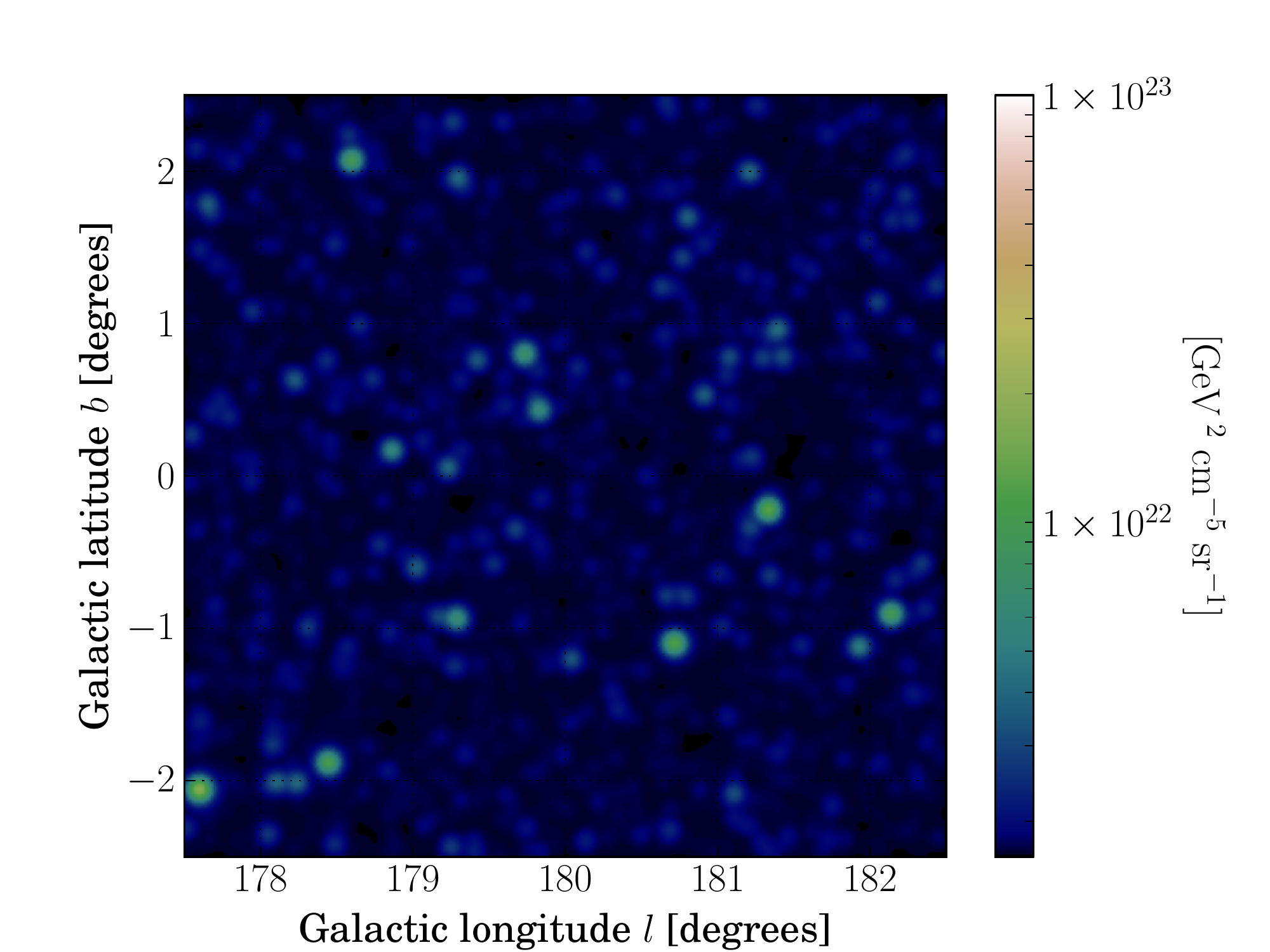}
\caption{{\em Top panel:} Example of a $5^{\circ} \times 5^{\circ}$ skymap towards the galactic anti-centre of the spherically symmetric halo, obtained from the skymap mode {\tt -g7} for a numeric resolution of $N_{\rm side} = 2^{13}$ (corresponds to a pixel diameter of $\sim 0.007^{\circ}$). The colour scale gives the $J$-values per steradian in case of annihilation. This skymap contains $64,{}763$ drawn substructures, computed with both a numeric precision and a relative error  $RE_{J_{clumps}}\leq 1\%$ 
within $2,{}5$ hours ($600\,MB$ RAM), using the physical parameters specified above. {\em Bottom panel:} Same skymap as on the top, but smoothed with a Gaussian beam of ${\rm FWHM} = 0.1^{\circ}$.}
\hspace{0.08\textwidth}
\label{fig:highrespatches}
\vspace{-0.5cm}
\end{figure}

For illustration of the new features available for the 2D-skymap mode  ({\tt -g5},  {\tt -g8} and {\tt -h4}, {\tt-h5}) of \clumpy{}, 
we provide example plots for Milky-Way like DM haloes in Figs.~\ref{fig:fullskymaps}, \ref{fig:highrespatches} and \ref{fig:powerspectra}. In these examples, the total density profile of the halo is parametrised by an Einasto profile with $r_{-2} = 15.14\,\rm{kpc}$, $\alpha_E = 0.17$ \cite{Fornasa2012} and a local dark matter density of $\rhosol= 0.4\,\rm{GeV\,cm}^{-3}$ at $\Rsol = 8.0\,\rm{kpc}$. For a typical demanded precision of $RE_{J_{clumps}}\leq 2\%$ (relative error), clumps down to masses of $\sim 1\,\Msol$ are drawn depending on the resolution, skymap size, \ldots (see Paper I \cite{2012CoPhC.183..656C} and especially Eq.~(28) in there for further details). The spatial distribution of subhaloes $\ud {\cal P}_V/\ud V$  is described by the {\tt kGAO\_SUB} profile,
\beq
\frac{N(< x)}{N_{200}} = \frac{(1 + ac)\,x^{\beta}}{1 +
   ac\,x^{\alpha}},\; x=\frac{r}{r_{200}}\;,
\label{eq:gao_profile}
\eeq
 with $ac = 11$, $\alpha=2$, $\beta= 3$ and $r_{200}\equiv \Rvir = 260\,\rm{kpc}$. Note that for such values, the outer slope of the subhaloes spatial distribution is $<2$, which does not allow the inclusion of more than one level of substructures as described in Sect.~\ref{subsubsec:extralevels}. The mass distribution $\ud {\cal P}_M/\ud M$ follows a power-law with index $\alpham = 1.9$, normalised by an abundance of $150$ subhaloes in the mass range $[10^{8}\,\Msol,\,10^{10}\,\Msol]$. The $c_{\rm vir}-R_{\rm vir}$ relationship is modelled by {\tt kSANCHEZ14\_200}, with a Gaussian log-norm scatter of $\sigma_c = 0.14$ around $\langle c_{\rm vir}\rangle$. A boost of the smooth $J$-factor is calculated taking into account substructures down to a minimal mass of $10^{-6}\,\Msol$. Only one substructure level is accounted for in these examples. For a more detailed description and listing of the input parameters entering these simulations, we refer to the documentation and the standard parameter file {\tt clumpy\_params.txt} delivered with the code. 

\autoref{fig:fullskymaps} presents full skymaps for an observer
located at $\Rsol = 8.0\,\rm{kpc}$, looking towards the centre of the
galactic halo ({\tt -g7} mode). The maps are given in terms of
differential $\gamma$-ray intensities at $4\,\rm{GeV}$ for
annihilation into $\chi\chi\rightarrow b\bar{b}$ channel and
$m_{\chi}=200\,\rm{GeV}$. We provide an example for a spherical and a
triaxial host halo shape. Flux and intensity maps both for {\gr} and
neutrinos can be generated from the the $J-$factor maps via the {\tt -f} option, or directly together with the $J-$factor calculation when specified in the input parameters. 

\autoref{fig:highrespatches} shows the spherical halo within a small patch of the sky computed with the highest possible resolution available in the \healpix{}-scheme, $N_{\rm side} = 2^{13}$, together with the same patch smoothed with an instrumental beam by the \healpix{} smoothing algorithm implemented in \clumpy{}. Here, the $J$-values are given per steradian; $J$-factors for arbitrary opening angles equal or larger than the grid resolution can be derived from this by integrating over the opening angle $\Delta \Omega$. For such high resolutions, the smoothing via spherical harmonics -- as done by the implementation in \clumpy{} -- is very computationally expensive ($\sim$ hours with $\geq 20\,GB$ RAM). For small FOV, smoothing via 2-dimensional Fourier transform in the tangential plane approximation is recommended instead; this feature is not yet implemented in \clumpy{}.

\begin{figure}[!t]
\centering
\includegraphics[width=0.49\columnwidth]{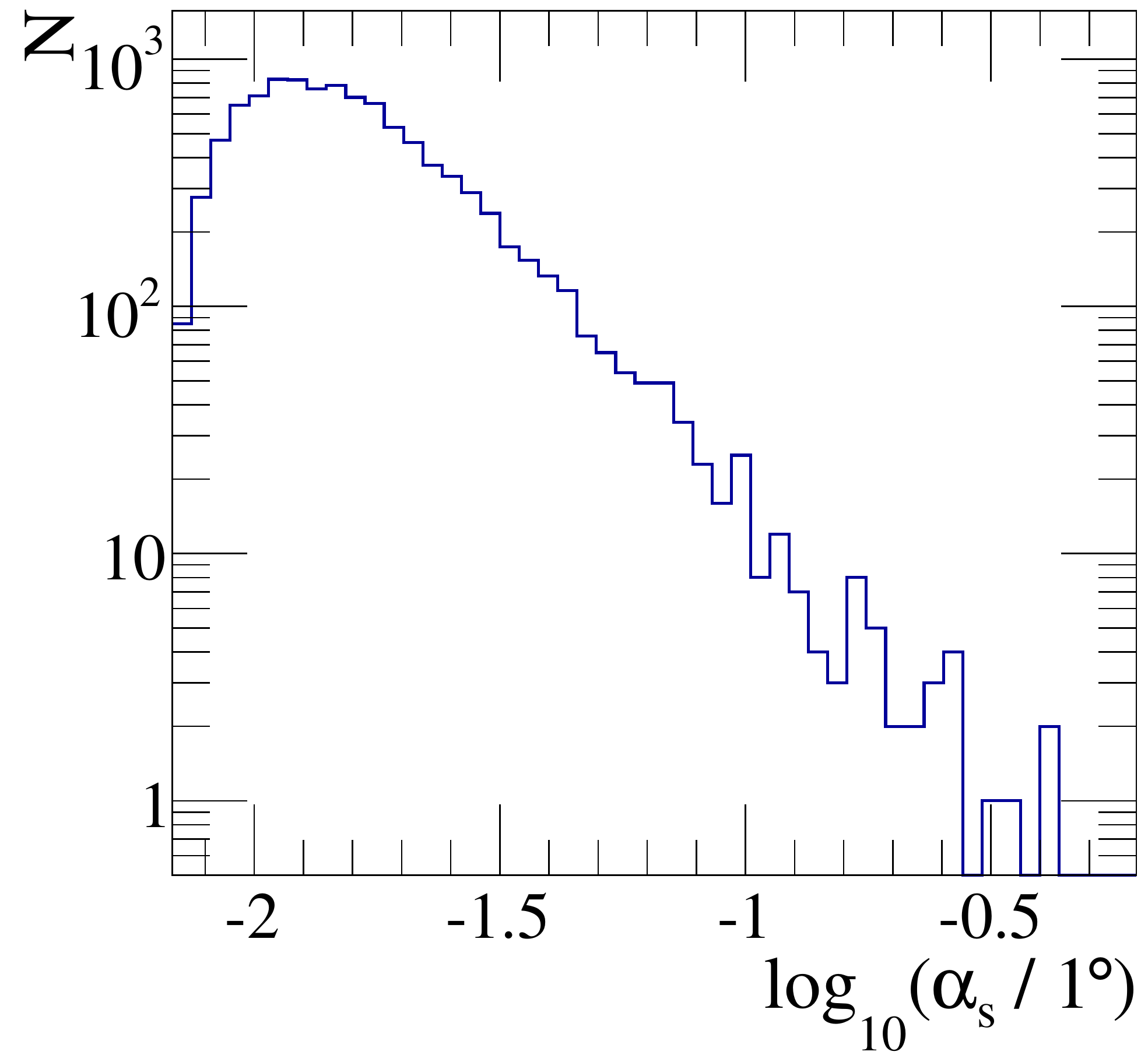}
\includegraphics[width=0.49\columnwidth]{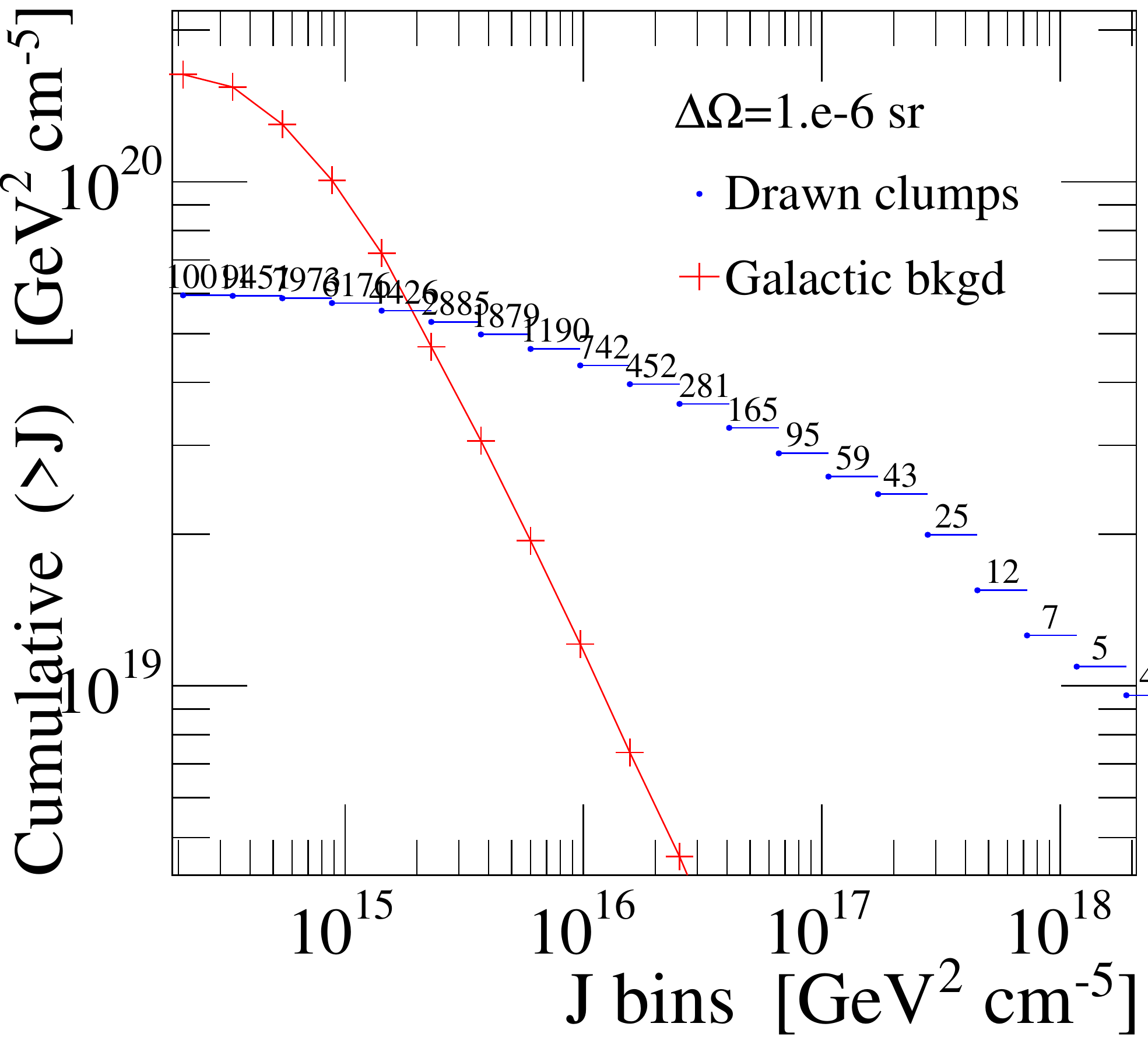}
\caption{Example of two population study plots obtained for galactic subhaloes towards the anti-centre
in a region of size $45^\circ\times45^\circ$ ({\tt ./bin/clumpy -g7 clumpy\_params.txt 180. 0. 45. 45. 1}). {\em Left panel:} Number of substructures drawn (above the DM continuum) as a function of their apparent size $\alpha_s=r_s/d$ ($r_s$ is the scale radius, and $d$ the distance from the observer). {\em Right panel:} cumulative of the signal of drawn subhaloes above a given $J$ (blue), and corresponding number $n$ of subhaloes contributing to this signal. The red curve corresponds to the cumulative from $n$ background regions.}
\hspace{0.08\textwidth}
\label{fig:popstudy}
\vspace{-0.5cm}
\end{figure}

\begin{figure}[!t]
\centering
\includegraphics[width=\columnwidth]{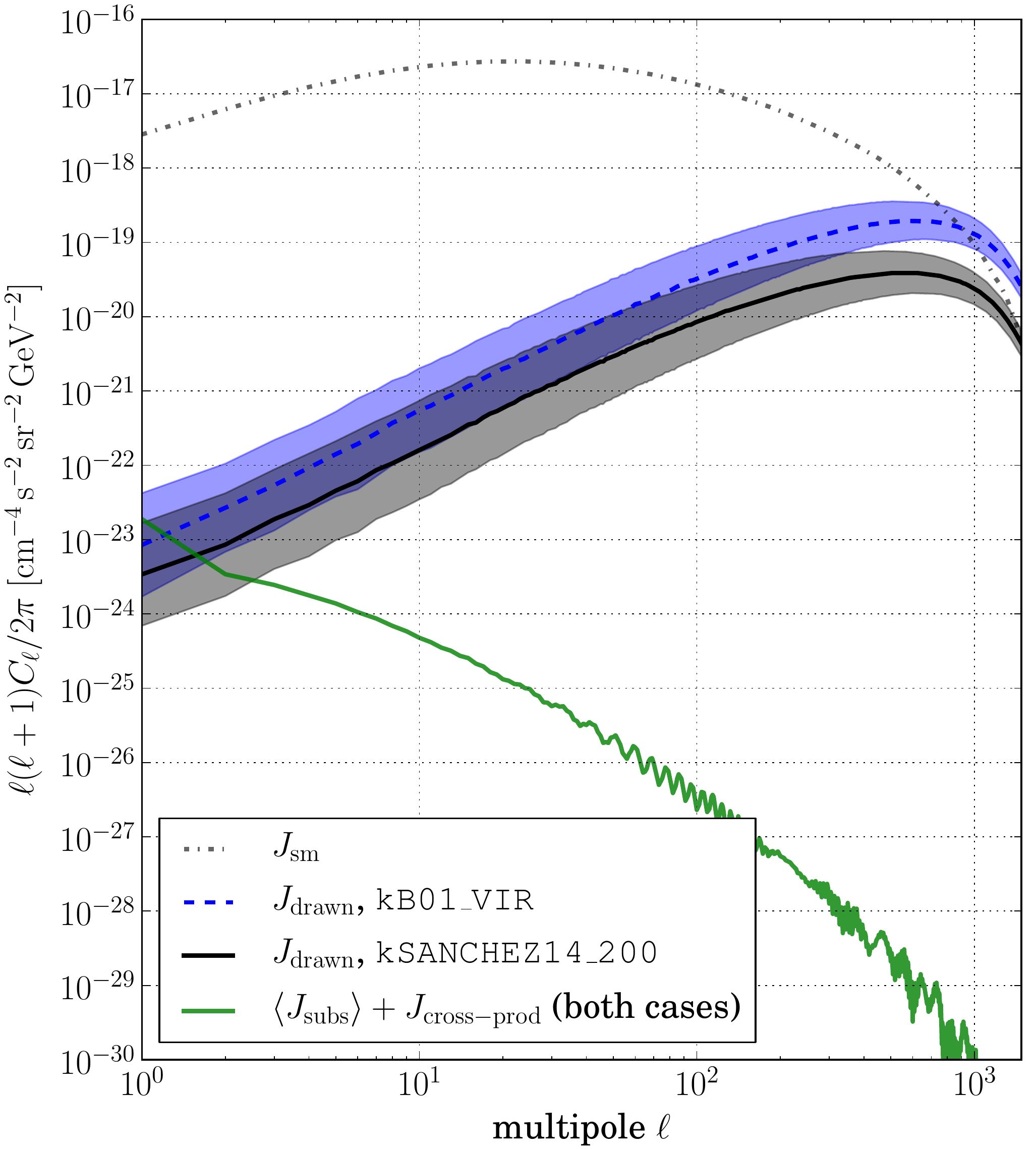}
\caption{Intensity APS for galactic substructures on the full sky for two different concentration models. For the particle physics term, $\gamma$-rays from annihilation at 4\;GeV for $m_\chi = 200$\;GeV and $\chi\chi \rightarrow b{\bar b}$ channel are assumed. The lines denote the mean value for 100 realisations of the same halo, the bands give the corresponding 1$\sigma$ containment range. The spectrum of the smooth halo $J_{\rm sm}$ (grey dotted-dashed line), and the
average contribution of the substructures ($\langle J_{\rm subs}\rangle + J_{\rm cross-prod}$, solid green line), are plotted for comparison.}
\label{fig:powerspectra}
\end{figure}

\autoref{fig:popstudy} shows two examples of plots generated by \clumpy{} when populations of DM haloes are available (e.g. from a user list or for clumps drawn in the Galaxy). For more illustrations, see the population study performed with \clumpy{} on clusters of galaxies in \cite{2012PhRvD..85f3517C,2012MNRAS.425..477N,2012A&A...547A..16M} and in \clumpy{} runs. Note that the list of subhaloes is stored in {\tt ASCII} files (extension {\tt .drawn}).

To better compare the results of \clumpy{} with results directly derived from the Aquarius \cite{2008MNRAS.391.1685S, Fornasa2012,Ando2013a}, Via Lactea II \cite{Diemand2008, 2015MNRAS.447..939L} and similar simulations of galactic haloes \cite{Calore2014}, \autoref{fig:powerspectra} displays the intensity angular power spectrum (APS) of the signal over the full sphere. The angular power spectrum $C_\ell$ of an intensity map $I(\vartheta, \varphi)$ is defined as
\beq
C_\ell = \frac{1}{2\ell + 1} \sum\limits_m |a_{\ell m}|^2 \;,
\label{eq:aps}
\eeq
with $a_{lm}$ the coefficients of the intensity map decomposed into spherical harmonics $Y_{lm}$,
\beq
I(\vartheta, \varphi) = \sum\limits_{\ell= 0}^{\ell_{max}} \sum\limits_{m= -\ell}^{m= +\ell} a_{\ell m}\, Y_{\ell m}(\vartheta, \varphi).
\label{eq:multipoledecomp}
\eeq

The filled bands correspond to the power spectrum of the drawn substructures, when considering a spherical host halo (using the same parameters as in Fig.~\ref{fig:fullskymaps}) but two different parametrisations of the $c_{\rm vir}-R_{\rm vir}$ relationship ({\tt kSANCHEZ14\_200} and {\tt kB01\_VIR}). Each spectrum was computed for $100$ different statistic realisations of the same galactic halo. The bands in \autoref{fig:powerspectra} indicate the $1\sigma$ containment ranges around the mean (assuming a log-normal variation of the $C_\ell$) of the cosmic variance of the simulated galactic halo. The total average contribution of the substructures ($\langle J_{\rm subs}\rangle + J_{\rm cross-prod}$, Eqs.~(\ref{eq:gal_meanJcl}) and~(\ref{eq:gal_Jcrossprod})) is given by the green solid line. Apart from the largest angular scales, the drawn substructures completely dominate over the APS of the averaged substructure contribution. The smooth signal (dotted-dashed line) dominates everywhere but for the smallest angular scales. The {\tt kSANCHEZ14\_200} model agrees well with the results presented in \cite{Fornasa2012}, where the resolved subhaloes from the Aquarius A-1 simulation \cite{2008MNRAS.391.1685S} are taken into account. These authors estimate the error on the APS, due to not resolving haloes below $10^{5}$\;M$_\odot$, to be $\lesssim 10\%$. In our approach, the full mass range for the subhaloes (down to $10^{-6}$\;M$_\odot$) is taken into account, within the $l_{\rm crit}$ criterion, to ensure the required user-defined accuracy on the total $J$-factor. Testing various values of the latter, we find that a $10\%$ accuracy requirement on the total $J$-factor translates into a $\sim 1\%$ error on the APS.

          %%%%%%%%%%%%%%%%%%%%%%%%%%%%%%%%%%
\subsection{Jeans: MCMC, PDF, CIs for $\rho$, $J$, $\sigma_p^2$\label{subsec:{cls}}}
Many useful quantities can be computed with \clumpy{} after a Jeans analysis, using the statistical options ({\tt -s}) of the \clumpy{} executable:
\begin{itemize}
  \item Figure \ref{fig:correl} shows a correlation plot made with the option {\tt -s2}. These results were obtained with the default options of the {\tt jeansMCMC} executable, i.e. using an MCMC analysis with the \great{} package. The $z$-axis of the 2D plots shows the value of $\log{\mathcal{L}}$.
\begin{figure}[!t]
\centering
\includegraphics[width=\columnwidth]{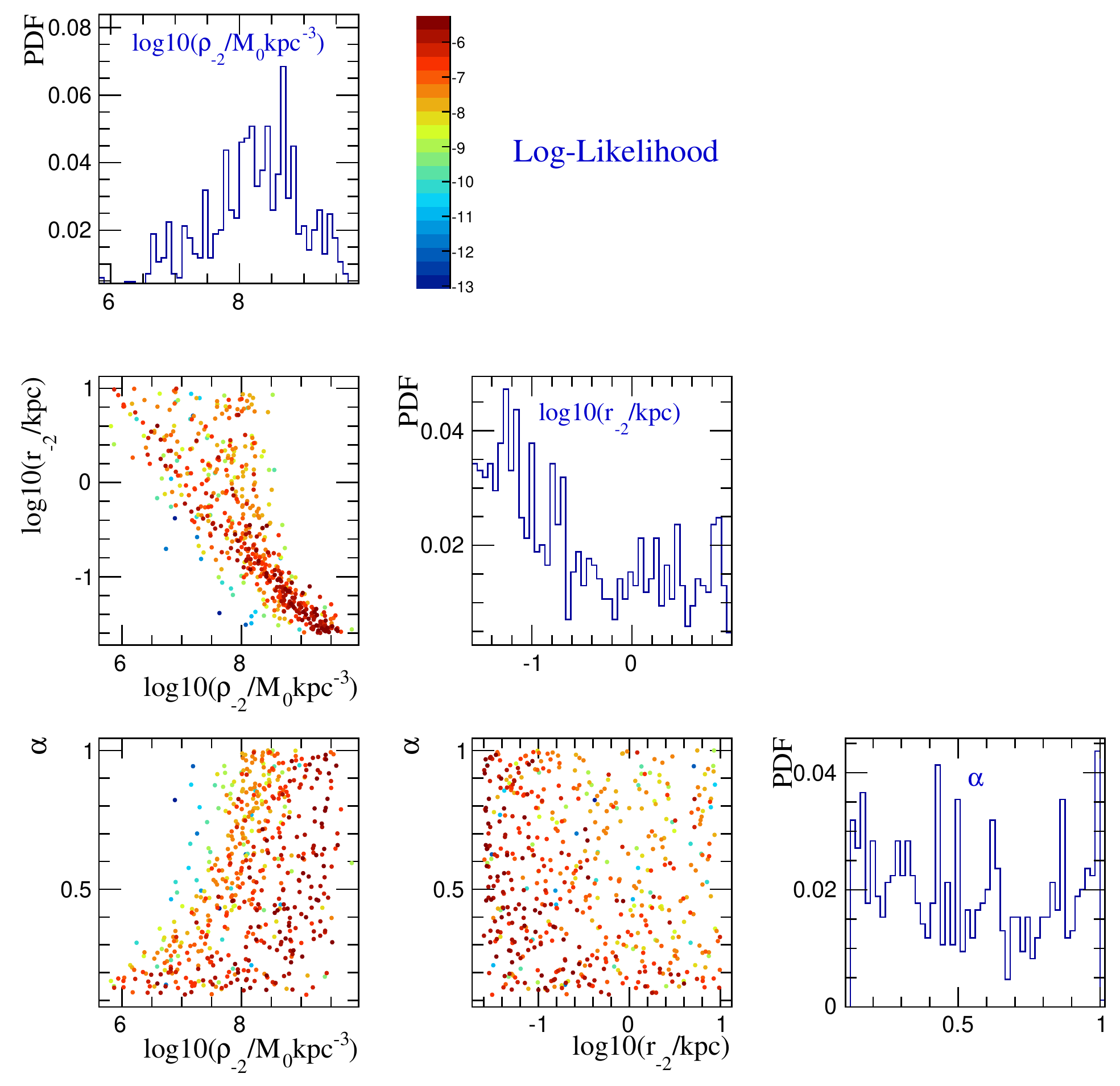}
\caption{Correlation plot obtained with the option {\tt -s2} of \clumpy{}. An MCMC analysis was run previously with the default options of the {\tt jeansMCMC} executable.}
\label{fig:correl}
\end{figure}
   \item Figure \ref{fig:CLs} displays two examples of quantities obtained with {\tt clumpy -s} after a Jeans analysis. The figure shows the median value and CIs on $\sigma_p(R)$ and $J(\alpha_{\rm int})$, obtained with the default options of {\tt clumpy -s8} and {\tt clumpy -s6} respectively. The statistical output file used was obtained with the default options of the {\tt jeansMCMC} executable and uses data of a mock ultra-faint dSph galaxy as provided with \clumpy{}. For outputs obtained on real data, we refer the reader to Figs. 1 and 2 of \cite{2015MNRAS.453..849B}\footnote{The Jeans analysis implemented in CLUMPY is based on a data-driven approach where as little prior from simulations as possible is used. In \cite{2015MNRAS.453..849B}, we have compared our results to that of the Fermi collaboration and did not find any systematic shift between the $J$-factors of the two analyses. However, our data-driven approach results in larger CIs for the ultra-faint dSph galaxies given the very few kinematic data available for these objects (see figure 6 of \cite{2015MNRAS.453..849B}).}.
\begin{figure}[!t]
\centering
\includegraphics[width=\columnwidth]{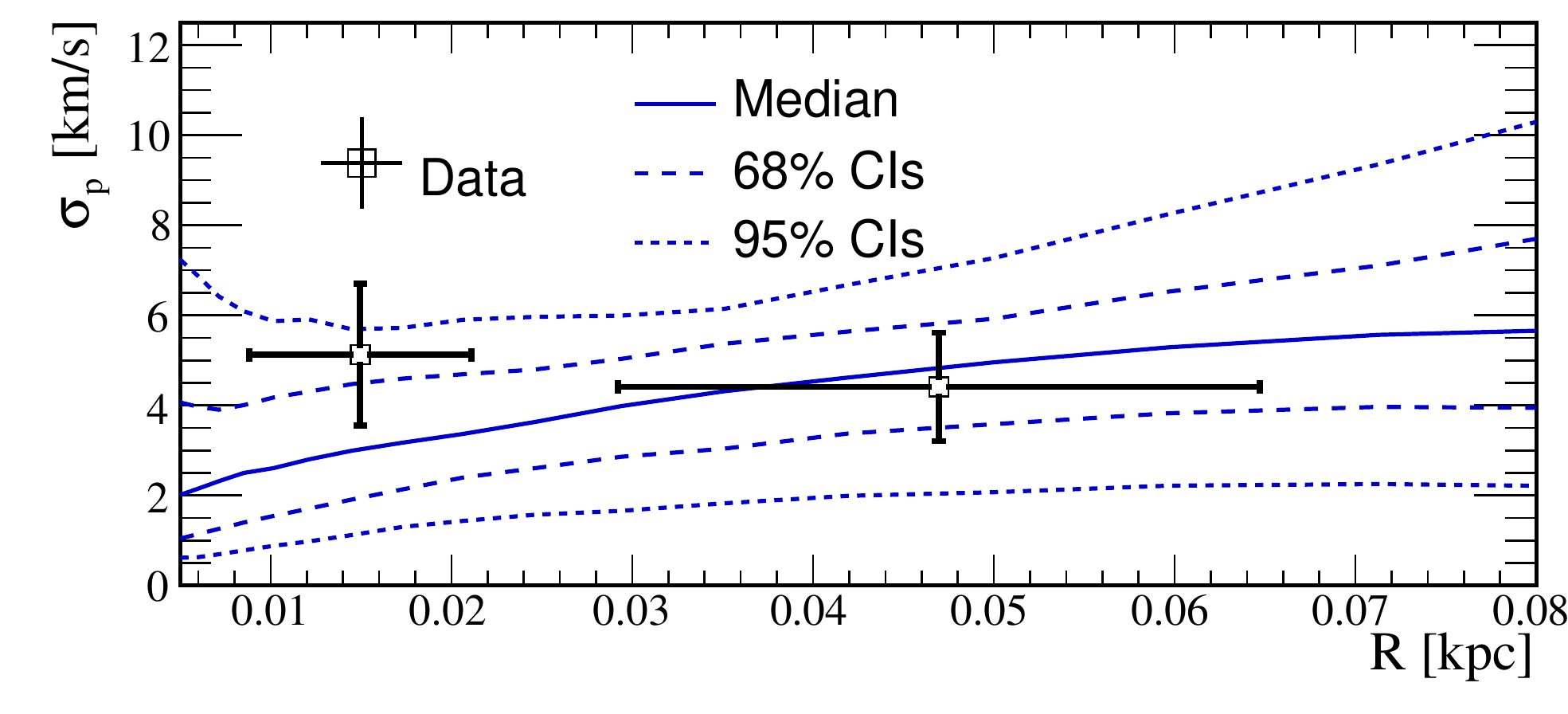}
\includegraphics[width=\columnwidth]{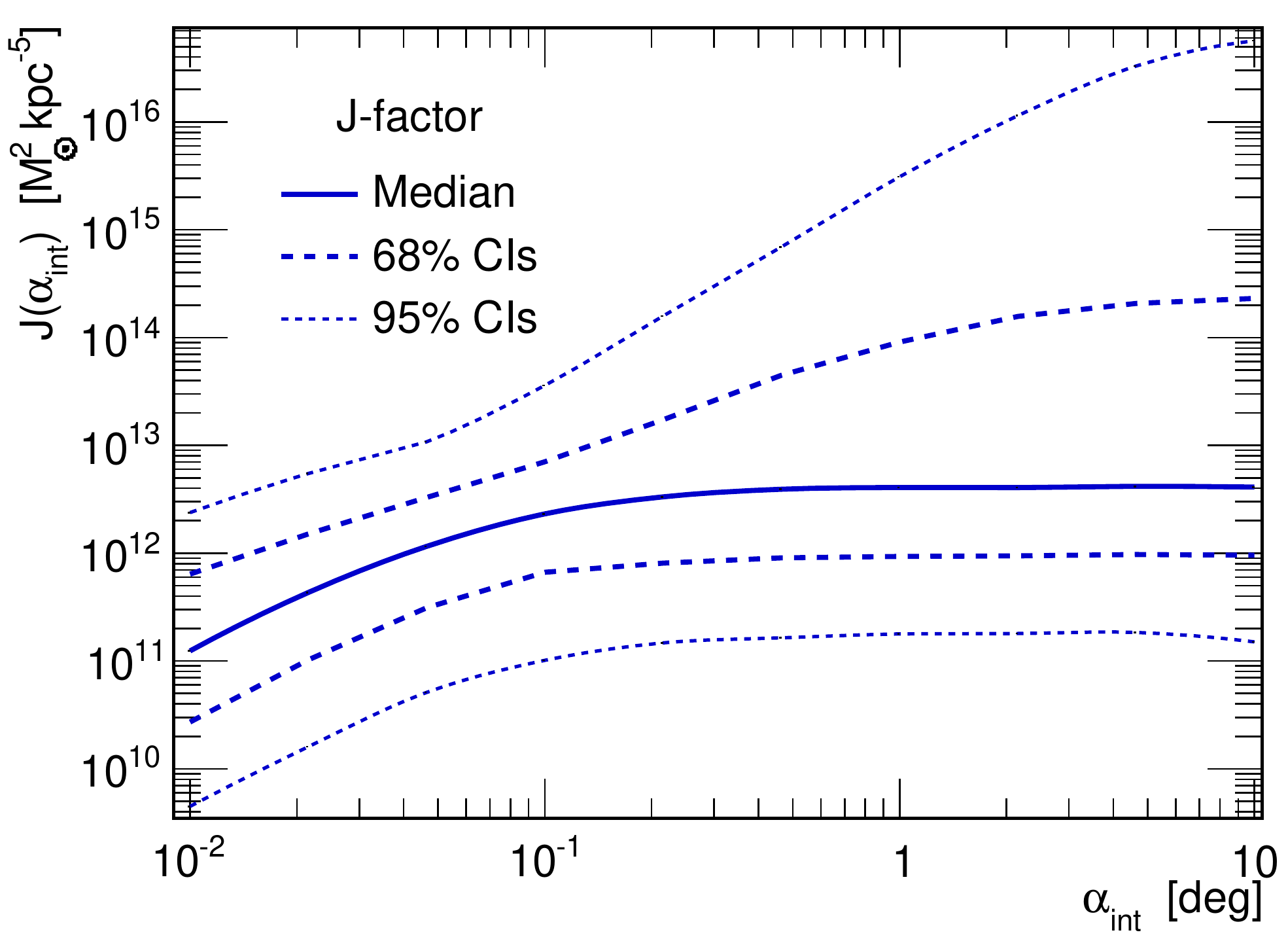}
\caption{Two examples of results obtained with the Jeans/MCMC analysis. The figure shows the median value and CIs on $\sigma_p(R)$ and $J(\alpha_{\rm int})$, obtained with the default options of {\tt clumpy -s8} and {\tt clumpy -s6} respectively.}
\label{fig:CLs}
\end{figure}
\end{itemize}

%%%%%%%%%%%%%%%%%%%%%%%%%%%%%%%%%%%%%%%%%%%%%%%%%%%%%%%%%%%%%%%%%%%%%%%%%%%%%%
%%%%%%%%%%%%%%%%%%%%%%%%%%%%%%%%%%%%%%%%%%%%%%%%%%%%%%%%%%%%%%%%%%%%%%%%%%%%%%
\section{Conclusions \label{sec:conclusion} }

The first version of the \clumpy{} code \citep{2012CoPhC.183..656C} was originally
developed to provide a fast, robust and versatile tool to compute
$J$ and $D$-factors for any halo, in a variety of configurations, and
with up to one level of substructures.
In this second release, several physically-motivated extensions have been
added, such as drawing halo concentrations from a distribution around
an average mass-concentration relation, including triaxiality
for DM haloes, allowing for several levels of substructures in the
boost calculation, and providing new analysis tools (e.g., $J$-factor ``population''
studies). Furthermore, the code now includes the PPPC4DMID calculation
for $\gamma$-ray and neutrinos yields, allowing the computation of
actual $\gamma$-ray and neutrino differential or integrated fluxes. 
Finally, the 2D mode (i.e. skymaps) is now handled in the \healpix{}
pixelisation scheme for an improved behaviour over large fractions of
the sky.

Along with these extensions, a new module was also developed
to perform Jeans analyses on a set of kinematic data (e.g. from dwarf spheroidal galaxies). This allows
the user to control the entire analysis chain, from the reconstruction of the DM density profile with the
kinematic data to the computation of the astrophysical factors. This is
done using either the \great{} MCMC engine or a simpler
$\chi^2$/bootstrap approach, depending on the user's choice. 

As for the first release, the code is fully documented and provides many examples.
Many options are available in command line executables, and several output formats
(\rootcern{} C++ based plots, \fits{} files and/or {\tt ASCII} and {\tt .root} files) are now
available, hopefully making \clumpy{} a user-friendly code for both
the particle physics and astrophysics communities.

\section*{Acknowledgements}
We thank M.~G.~Walker for many useful discussions and cross-checks in the
development of the Jeans module, L.~Derome and A.~Putze for
helping to interface \clumpy{} with \great{}, and G.~Maier and L.~Gerard for 
the proofreading and valuable comments on the manuscript. E.~N. would like to thank M.~Cirelli. The work of M.H.
was supported by the Research Training Group 1504 of the German Research Foundation (DFG).
This work has been supported by the ``Investissements d'avenir, Labex
ENIGMASS", and by the French ANR, Project DMAstro-LHC,
ANR-12-BS05-0006.

%%%%%%%%%%%%%%%%%%%%%%%%%%%%%%%%%%%%%%%%%%%%%%%%%%%%%%%%%%%%%%%%%%%%%%%%%%%%%%
%%%%%%%%%%%%%%%%%%%%%%%%%%%%%%%%%%%%%%%%%%%%%%%%%%%%%%%%%%%%%%%%%%%%%%%%%%%%%%
 %\appendix
%\section{Geometry and change of coordinates \label{app:geom}}

%\bibliography{clumpy2}
%\bibliographystyle{cpc}

\end{document}